\documentclass[preprint,12pt]{elsarticle}

\usepackage{amssymb}
\usepackage{amsmath}

\usepackage{xspace}
\usepackage{tikz}
\usepackage{graphicx}
\usepackage{color}
\usepackage{amssymb}
\usepackage{amsmath}
\usepackage{amsfonts}
\usetikzlibrary{decorations.pathreplacing,positioning, arrows.meta}
\usepackage{hyperref}
\usepackage{mathtools}

\usepackage{algorithm}
\usepackage{algpseudocode}
\usepackage{caption}
\usepackage{tocloft}
\usepackage{makecell}
\usepackage{ragged2e}
\usepackage{lscape}

\journal{Robotics and Autonomous Systems}

\newcommand{\cpp}{$\text{C}^{++}$}

\newcommand{\ttt}{\mathtt{true}}
\newcommand{\fff}{\mathtt{false}}

\newcommand{\tx}{\mathbf{X}}
\newcommand{\tf}{\mathbf{F}}
\newcommand{\tg}{\mathbf{G}}
\newcommand{\tu}{\mathbf{U}}

\newcommand{\trans}{\longrightarrow}
\newcommand{\fun}{\longrightarrow}

\newcommand{\buchi}{B{\"u}chi\xspace}

\usepackage{listings}
\lstdefinelanguage{scsl}{
  morekeywords={
    global, const, type, object, scenario, parameters, precondition, spec, end, constraint,
    interface, from, to, initact, cndact, elementary, of, instance, function, collaboration, for,
    delete, create, systemtest, enum, schedule, cycletime, if, then, endif, else
  },
  morecomment=[l]{--}}
\lstdefinestyle{scsl}{
  frame=no,
  mathescape=true,
  breaklines=true,
  basicstyle=\footnotesize\color{black},
  keywordstyle=\sffamily\color{black}\bfseries,
  identifierstyle=\ttfamily\color{blue}\slshape,
  stringstyle=\ttfamily,
  commentstyle=\ttfamily\color{gray},
  emphstyle=\ttfamily\color{blue},
  tabsize=4,
  numbers=left,
  numberstyle=\tiny,
  stepnumber=1,
  numbersep=5pt,
  columns=flexible}

\lstdefinestyle{bash}{
	frame=no,
	mathescape=true,
	breaklines=true,
	basicstyle=\ttfamily\footnotesize\color{black},
	keywordstyle=\ttfamily\color{black},
	identifierstyle=\ttfamily\color{black},
	stringstyle=\ttfamily\color{black},
	commentstyle=\ttfamily\color{gray},
	emphstyle=\ttfamily\color{black},
	tabsize=4,
	numbers=left,
	numberstyle=\ttfamily\tiny,
	stepnumber=1,
	numbersep=5pt,
	columns=flexible
}

\newcommand{\inp}{\mathbf{in}\ }
\newcommand{\oup}{\mathbf{out}\ }

\newcommand{\tsc}{\hat{\mathtt{t}}}

\newcommand{\frm}{\underline{\mathrm{frame}}}

\newcommand{\const}{\mathtt{const}~}

\newcommand{\sactive}{\text{\tt active}}

\newcommand{\tend}{\text{\tt EoT}}

\newcommand{\rnneg}{\mathbb{R}_{\ge 0}}

\newcommand{\OEQ}{{\tt OE}\xspace}
\newcommand{\OEH}{{\tt OEH}\xspace}
\newcommand{\SIM}{{\tt SIM}\xspace}
\newcommand{\PER}{{\tt PER}\xspace}

\newcommand{\chgc}[1]{\text{\ttb when}\ (~#1~)}

\newcommand{\location}{\text{\tt Location}}
\newcommand{\status}{\text{\tt Status}}
\newcommand{\stuck}{\text{\tt stuck}}
\newcommand{\fault}{\text{\tt fault}}
\newcommand{\atdst}{\text{\tt atDst}}
\newcommand{\itemloaded}{\text{\tt itemLoaded}}
\newcommand{\loadedret}{\text{\tt returnedWithItem}}
\newcommand{\returned}{\text{\tt returned}}
\newcommand{\initial}{\text{\tt initial}}
\newcommand{\approaching}{\text{\tt approaching}}
\newcommand{\returning}{\text{\tt returning}}
\newcommand{\returningwi}{\text{\tt returningWithItem}}

\newcommand{\rover}{\text{\sl Rover}}

\newcommand{\rcmds}{\text{\tt RoverCommands}}
\newcommand{\gotodst}{\text{\tt goToDst}}
\newcommand{\returntodst}{\text{\tt returnToDst}}
\newcommand{\pickup}{\text{\tt pickUpItem}}

\newcommand{\narea}{\text{\tt numZones}}
\newcommand{\nga}{\text{\tt exclusionZone}}
\newcommand{\area}{\text{\tt Zone}}
\newcommand{\tdst}{t_\text{atDst}}
\newcommand{\tcend}{t_\text{end}}
\newcommand{\numr}{\text{\tt numRovers}}

\newcommand{\innga}{\text{\tt inExclusionZone}}
\newcommand{\isclose}{\text{\tt isCloseTo}}

\newcommand{\scnull}{\text{\tt null}}
\newcommand{\startPos}{\text{\tt startPos}}
\newcommand{\returnDst}{\text{\tt returnDst}}
\newcommand{\allIds}{\text{\tt allIds}}
\newcommand{\itemId}{\text{\tt ItemId}}
\newcommand{\targetDst}{\text{\tt targetDst}}

\newcommand{\dom}{\mathsf{dom}~}

\newcommand{\ttb}{\ttfamily\fontseries{b}\selectfont}

\newcounter{examplectr}

\floatstyle{plain} 
\newfloat{myscenario}{htb}{lob}
\floatname{myscenario}{\bf Scenario}

\DeclareCaptionFormat{custom}{#1#2#3}
\newfloat{myglobalspec}{htb}{lob}
\floatname{myglobalspec}{Global Specification}

\newlistof{myscenariolist}{myscenariolistcnt}{List of Scenarios}
\newlistof{myglobalspeclist}{myglobalspeclistcnt}{List of Global Specifications}

\newcommand{\addtomyscenariolist}[1]{%
	\refstepcounter{myscenariolist}
	\addcontentsline{myscenariolistcnt}{myscenariolist}{\protect\numberline{\themyscenariolist}#1}\par
}
\newcommand{\addtomyglobalspeclist}[1]{%
	\refstepcounter{myglobalspeclist}
	\addcontentsline{myglobalspeclistcnt}{myglobalspeclist}{\protect\numberline{\themyglobalspeclist}#1}\par
}

\begin{document}

\begin{frontmatter}



\title{Scenario-based System Testing for Distributed Robotics Applications}


\author[label1]{Jan Peleska}
\author[label2]{Felix Brüning}
\author[label1]{Wen-Ling Huang}
\author[label3]{Anne E. Haxthausen}

\affiliation[label1]{organization={University of Bremen},
	addressline={Bibliothekstraße 1}, 
	city={Bremen},
	postcode={28359}, 
	state={Bremen},
	country={Germany}}
	
\affiliation[label2]{organization={Verified Systems International GmbH},
	addressline={Am Fallturm 1}, 
	city={Bremen},
	postcode={28359}, 
	state={Bremen},
	country={Germany}}
	
\affiliation[label3]{organization={DTU Compute, Technical University of Denmark},
	addressline={Building 322}, 
	city={Lyngby},
	postcode={2800}, 
	state={Lyngby},
	country={Denmark}}

\begin{abstract}
We present the \textbf{SC}enario \textbf{S}pecification \textbf{L}anguage (SCSL) for automated generation and execution of system-level tests. SCSL targets complex distributed systems (e.g., collaborating autonomous robots) where classical model-based testing becomes impractical because (1) the overall system complexity is too high for a single monolithic model, (2) test behaviour cannot be fully precomputed due to substantial nondeterminism in the distributed system under test (SUT), and (3) the SUT configuration may change dynamically at runtime.
Challenge (1) is addressed by scenarios: each scenario specifies test-specific expected SUT behaviour and/or stimuli to be applied during execution. Complex system tests are composed from elementary scenarios using sequential and parallel composition. To address (2), the SCSL tool platform supports online (on-the-fly) testing, selecting and executing test steps during runtime. For (3), SCSL provides a collaboration construct that supports dynamic reconfiguration: removing unavailable components, registering newly joining components, and rewiring interfaces during test execution.
We illustrate the syntax and semantics of SCSL using a system-test example in which robots perform a salvage mission, and we use an automatically generated test execution to demonstrate the concepts supported by our prototype tool platform.
\end{abstract}


\begin{keyword}
System testing
\sep
Scenario-based testing
\sep
Collaborating robots
\sep
Linear Temporal Logic




\end{keyword}

\end{frontmatter}



\section{Introduction}\label{sec:intro}

\subsection{Objectives}

The objective of this article  is to present the \textbf{SC}enario \textbf{S}pecification \textbf{L}anguage (SCSL) for defining and executing functional system tests for complex cyber-physical systems (CPS). The presentation uses a typical example of a system test for several robots collaborating in a joint mission. 
When comparing ``classical'' model-based system testing (MBT) to the scenario-based approach, the crucial differences are as follows.
\begin{itemize}
\item MBT aims to capture all admissible system behaviours within a comprehensive system model. From this model, test cases, test data, and test procedures are derived to achieve a certain level of coverage.  This model coverage may also be related to requirements coverage, if the underlying formalism allows for tracing model elements to requirements and vice versa~\cite{DBLP:conf/isola/0001BH18}.

\item Scenarios are more restrictive behavioural models, each scenario describing a relevant behavioural aspect of the system, for example, a traffic scenario for an autonomous vehicle. Each relevant system behaviour should be covered by at least one scenario. The collection of all scenarios, referred to as the \textit{scenario library}, serves as a structured specification of system behaviour for testing purposes. A separate validation activity not discussed in this paper  ensures that all relevant behaviours are covered by the library~\cite{DBLP:conf/itsc/HauerSHP19}.

\item Typically, scenarios types are parameterised, so that they can be instantiated in different system test situations with different parameter values. A scenario instance is a concrete behavioural specification, where all parameters have been associated with concrete values (so-called \emph{actual parameters}). 

\item The objects under test -- in the example presented below, the collaborating robots and a command centre supervising the mission -- are input parameters to the test scenarios, and their expected behaviour is specified in the scenario, since only the scenario-specific behavioural aspects need to be described.

\item A scenario-based system test consists of a set of scenario instances that define a meaningful end-to-end system behaviour in alignment with specific test objectives.
\end{itemize}

Consequently, scenarios are easier to specify than the reference models needed in MBT, where the expected behaviour of a system under test (SUT) needs to be modelled for {\it all} operational situations.

Apart from their complexity, the distributed CPS that have influenced the design of the SCSL typically pose two further challenges: The local autonomy of the SUT components induce a very high degree of nondeterminism. Consequently, pre-calculated linear test steps will often be unsuitable to reach the system test objectives.  Therefore, the underlying tool platform needs to support \emph{online} testing, where the consecutive test steps are calculated at runtime, based on the currently observable state of the SUT. Note that this capability just depends on the underlying tool implementation, it is possible both for MBT (see Larsen et al.~\cite{10.1007/978-3-540-31848-4_6}) and scenario-based testing.

The final challenge consists in the fact that complex distributed CPS may change their configuration at runtime: robots, for example, may autonomously decide to leave or enter the SUT configuration, they may no longer be available due to some component failure, and interfaces may be ``rewired'' at runtime. Therefore, modelling formalisms  only able to describe static process networks~\cite{DBLP:journals/dc/BrookesR91} are unsuitable for describing (comprehensive models or) scenarios for testing these CPS.

\subsection{Test Configurations}

\begin{figure}[H]
\begin{center}
\includegraphics[width=\textwidth]{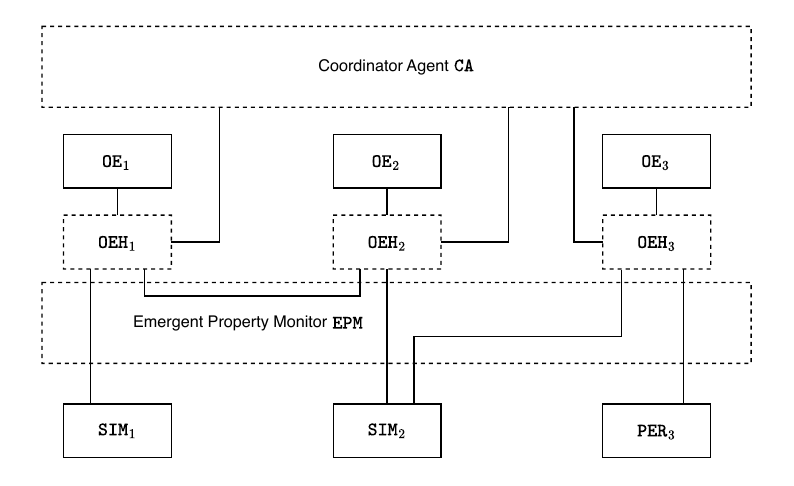}
\end{center}
\caption{System test configuration.}
\label{fig:systestconfig}
\end{figure}

Typical test configurations for the scenario-based approach advocated by the authors will be variants of the one depicted in Figure~\ref{fig:systestconfig}. The CPS under test is a distributed system where control computers communicate with each other and may act on electro-mechanical peripherals. In a system test configuration, some of these control computers will be present as \emph{original equipment \OEQ}, while others are replaced by \emph{simulations \SIM}. Hardware peripherals could also be replaced by other simulations, or they could be present as original \emph{peripherals \PER}. The solid lines between boxes each represent one or more interfaces for data exchange. Interfaces (to be introduced in Section~\ref{sec:systestconfig}) realise uni-directional data flows between components participating in a system test.

The objective of system tests is to check the   functional behaviour of all original equipment   and peripherals, as well as their interactions. The ``local'' behaviour of original equipment is monitored by \emph{original equipment harnesses \OEH}. These harnesses run test-specific \emph{elementary scenarios} specifying the expected behaviour of their \OEQ  during the system test, as well as the stimulations to be provided for some \OEQ input interfaces, as far as these are not provided by other \OEQ or by simulations or peripherals.

The succession of elementary scenarios to be executed by each \OEH is coordinated by a \emph{coordinator agent {\tt CA}}. The {\tt CA} receives status information  about ongoing elementary scenario executions performed by each \OEH and provides further elementary scenarios to be scheduled by the associated \OEH. The collaboration between all components of the CPS will result in certain \emph{emergent properties}~\cite{DBLP:journals/csur/NielsenLFWP15} achieved by the overall system. These can be checked by an \emph{emergent property monitor {\tt EPM}} observing all communications between CPS sub-components and its operational environment (typically replaced by simulations during a system test).

\subsection{Contributions}

\paragraph{Previous work by the authors}
This article is based on the authors' conference contribution to TAROS2025~\cite{DBLP:conf/taros/PeleskaBHH25}.
There, the SCSL syntax and its semantics have been introduced in   abbreviated form. A robotics system test where all scenarios acted as observers (i.e.~test oracles) had been used to illustrate SCSL usage.
While other existing scenario modelling languages are frequently domain-specific, the SCSL has been expressly designed to allow for applications in \textit{any} domain: Domain-specific details (e.g.~physical laws) can be encapsulated in separate elementary scenarios that are collected in a re-usable library. Another distinguising feature of the SCSL is that elementary scenarios may combine declarative and imperative specification styles: 
Linear Temporal Logic (LTL) allows for declarative specifications, and a guarded action language allows for imperative specification.
From our experience, this facilitates behavioural modelling in a considerable way.
\begin{itemize}
\item Declarative specifications are good for including all acceptable behaviours without having to enumerate them algorithmically. On the downside, it may be quite hard to produce declarative specifications that capture all acceptable behaviours without inadvertently including some unwanted behaviours. 

\item Imperative   specifications, when designed with a focus on simple readability, may exclude alternative acceptable solutions.
\end{itemize}
A further distinguishing SCSL feature is that  its semantics takes explicitly into account that typical controllers in a CPS run in \emph{execution cycles} of constant duration. New inputs are read and outputs are written at the beginning of each cycle.   Therefore, they no longer behave like perfect physical models whose differential equations specify piecewise smooth   evolutions for each real-valued observable (such as location or velocity). Instead, controllers produce variable updates at each cycle that should be close approximations of the discretised differential equations of the underlying physical models. Our position (which is also shared by other authors working in the field of CPS~\cite{DBLP:conf/iceccs/CavalcantiH23}) is that formalisms using  a semantics close to physical models (like, for example, the 
Multi-Lane Spatial Logic MLSL~\cite{DBLP:conf/icfem/HilscherLOR11}) should only be used to describe the idealised physical view 
of a CPS, while modelling formalisms used to derive test cases for the implemented system should take into account the cyclic nature of the controllers involved. 

The challenges posed by SUTs with dynamically changing configurations have been solved by means of a collaboration specification which is part of an SCSL system test specification and my be modified during the test execution.

\paragraph{New contributions in this article}
For this article, we have extended our previous work as follows.
\begin{enumerate}
\item The elementary scenario syntax has been extended by auxiliary $\frm$-variables. These make it easy to specify whether an elementary scenario should act as an observer or a stimulator, influencing SUT reaction: An orginal equipment harness running an elementary scenario instance may change object parameters that are registered in the frame    to ensure that the LTL specifications are fulfilled. If the frame is empty throughout the test execution, the scenario instance acts as an observer, and a violation of an LTL specifications is reported as a failure of the test execution.

\item The system test example has been modified to allow for both observer scenario instances and simulated equipment  instances.

\item Auxiliary object have been introduced to facilitate inter-scenario communication at runtime.

\item The section on related work has been updated to take latest developments into account.

\item The behavioural semantics of scenarios is now explained in full detail in an appendix.  

\item A new section presents a system test example with automatically generated data. This is also used to illustrate the basic functionality of the underlying prototype tool platform.
\end{enumerate}

While contributions 1 --- 4 are minor (but in case of 1,2,3 quite effective) extensions, contributions 4 and 6 are substantial.

\subsection{Overview}

 In Section~\ref{sec:systest}, the system test used as an example for introducing the SCSL is described in natural language.
 The SCSL objects and scenarios involved, as well as the resulting system test configuration are presented in Section~\ref{sec:scenariosystest}. In this section, the  SCSL syntax and its intuitive semantics are explained as well.
A system test exeacmple with explanations on the SCSL prototype platform is discussed in Section~\ref{sec:tools}. In Section~\ref{sec:related}, related work is discussed. Section~\ref{sec:conc} contains a conclusion and discusses future work. In \ref{sec:scslsem}, we present a more formal comprehensive explanation of SCSL behavioural semantics.

\section{A System Test   for Verifying Distributed Collaborating Robots}\label{sec:systest}

\paragraph{Mission}
The robots' mission is to salvage goods   from a known location in a hazardous area, where a salvage by humans is deemed to be impossible, due to various adverse environmental  conditions. 

\paragraph{Goods to be salvaged}
The goods consist of $m >0$  items (like boxes or barrels) that can be picked up by a robot arm and loaded onto a transport-capable robot.  Each item has a unique id, which can be spotted by the robots.\footnote{Typical identification techniques could be QR codes or RFID tags}

\paragraph{Terrain}
We assume that the terrain where the goods are located  has several exclusion zones (such as cliff edges), so that the robots cannot reach their destination on a straight path from their starting points.

\begin{figure}[H]
\begin{center}
\includegraphics[width=.7\textwidth]{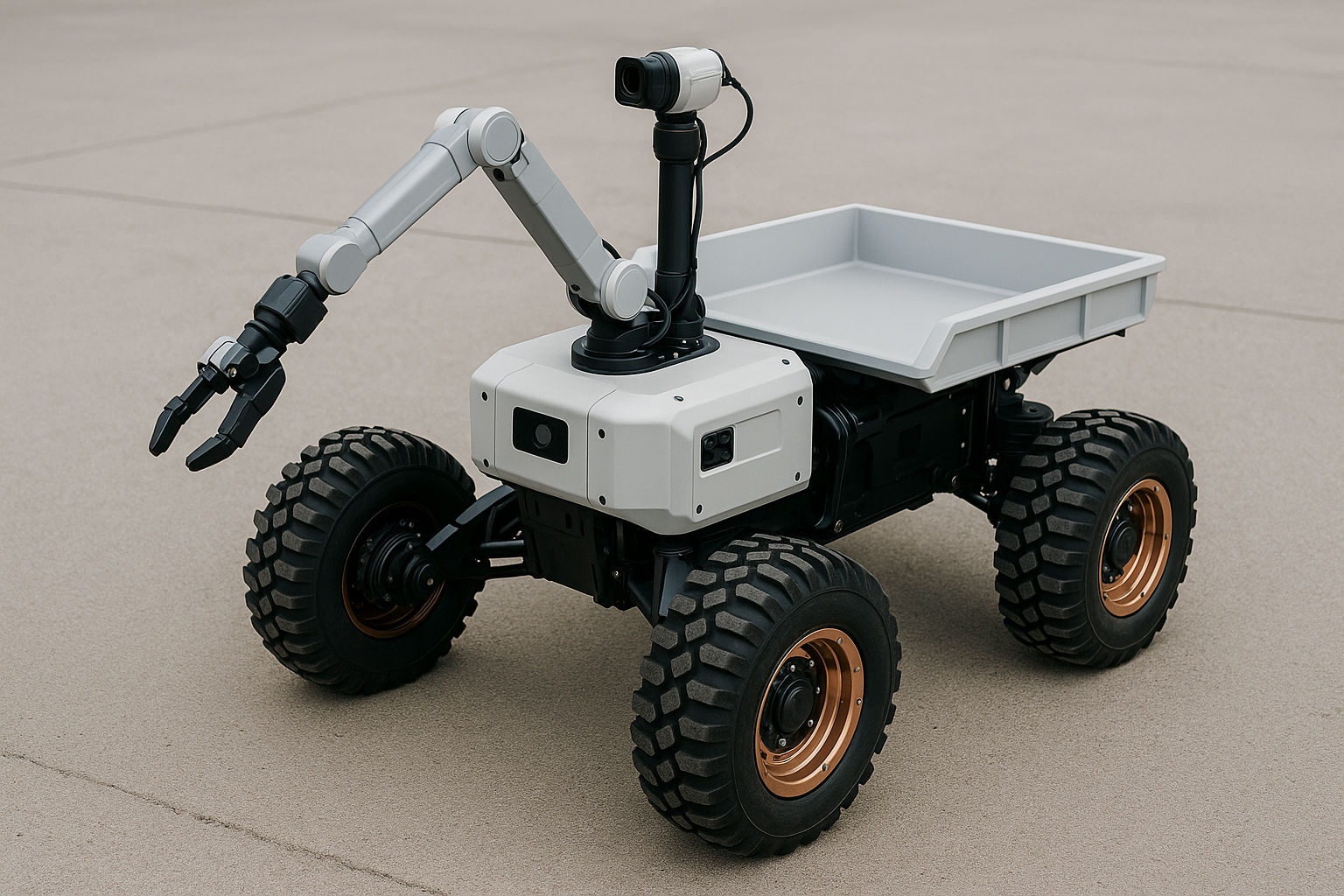}
\end{center}
\caption{Rover-type robot.}
\label{fig:rover}
\end{figure}

\paragraph{Robots}
There are $n$ robots   of the \emph{rover} type (see Fig.~\ref{fig:rover}): they move on wheels that are suitable for the terrain. Each rover is identically equipped and has the same capabilities: 
(1)~A robot arm to pick up an item.  
(2)~A cargo bed where an item can be placed.
(2)~Sensors allowing to detect the boundaries of exclusion zones.
(3)~A GPS sensor to determine its location.
(4)~Autonomous moving capabilities to circumvent exclusion zones and find a path to a given destination, 
if such a path exists from their actual location.
(5)~Cameras, image recognition functions, RFID readers to detect and pick up an item and place it on a robot's flatbed.
(6)~Communication equipment to send status information to a command centre and receive commands from there. 
(7)~Controllers to process the commands received.

\paragraph{Mission planning}
As the optimal path to the goods is unknown, rovers are deployed from different starting points, but they all get the same  destination coordinates where all items to be salvaged are located. It is expected that some rovers will get stuck on their path and be unable to reach the destination. Others may have hardware faults, and some might be  destroyed by inadvertently entering an exclusion zone. Therefore, $n$ rovers are initially sent on the mission, while only $m < n$ rovers are needed to pick up and return the salvaged goods. The first $m$ 
  to reach the destination will be commanded to pick up the goods. To this end, the command centre sends different item ids to the rovers, so that they all approach and load different items.  As soon as a rover has managed to load an item onto its flatbed, it is commanded to return to a specific location. If a rover becomes unavailable, either due to entering an exclusion zone or experiencing a failure, the command centre may reassign the corresponding task to another rover that has already returned with an item, instructing it to collect a new item that was originally assigned to the now unavailable rover.
All rovers that get stuck or fail during their mission are commanded to return (even if it is uncertain whether they will be able to return at all). The same holds for   operative rovers that are no longer needed since all items to be salvaged have been picked up by other robots.
Return locations may vary between rovers, especially for those that failed to reach the destination and can exit the hazardous area more quickly via alternative paths. 

\paragraph{Pass criterion}
The system test passes if at least $m \le n$ rovers arrive before time $\tdst$ (arrival deadline) and 
at least $k \le m$ rovers return  carrying salvaged items before time $\tcend > \tdst$ (mission completion deadline).

\paragraph{System test configuration}
For the system  test considered here, it is assumed that all robots are available as original equipment, but that the command centre should be simulated by the testing environment.

\section{Scenario-based System Test Specification in SCSL}\label{sec:scenariosystest} 

\subsection{Overview}

A complete SCSL system test specification consists of (1)~elementary type specifications, such as enumerations, (2)~composite data type specifications, similar to mathematical data  types in Z or VDM~\cite{spivey1992znotation,jonesVDM}, (3)~global constants and function definitions, (4)~object type specifications for the types of components participating in a system test, (5)~elementary scenario type specifications, and (6)~a system test configuration.

A system test configuration references all the participants in the  test execution, as well as their means of interaction, and the test execution schedule. The participants are instances of object type specifications. Each object  represents original equipment, peripherals, or simulations thereof. Auxiliary objects may be used as data containers storing information to be communicated between elementary scneario instances. The means of interaction between objects are provided by \emph{interfaces}. Objects and interfaces are comprised in a \emph{collaboration}\footnote{The term collaboration is used here in a generic sense to denote the set of participants and their interconnections; unlike UML~\cite{uml_2_5} collaborations, SCSL collaborations are instance-based and dynamically reconfigurable.} which is the first part of a system test configuration.

The \emph{execution schedule} (second part of the system test configuration) shows how original equipment \textit{should} evolve (i.e.~expected/checked at runtime) and how simulations \textit{will} evolve (i.e.~enforced by the test environment) over time during the system test execution.
An elementary scenario refers to one or more objects and specifies their required behaviour for that particular test situation. Crucially, SCSL follows a \emph{scenario-centric style}: object types provide interfaces and state/data structure, while the situation-dependent behaviour relevant to a test step is specified in the scenarios that involve the objects (and their parameters).
\begin{quote}
Instead of specifying the full behavioural model of an object type \textit{inside the object type}, the behaviour required in a given test situation is specified \textit{in an elementary scenario} that refers to the object and instantiates its parameters.
\end{quote}

In the terms introduced in Section~\ref{sec:intro}, elementary scenarios runs on an original equipment harness (\OEH).
Note that system test schedules usually involve more than one elementary scenario, since this type of test aims at verifying end-to-end functionality, as perceived by end users of the system. In  the system test of the robots described in this paper, for example,  separate elementary scenarios are used to describe robots approaching the salvage destination, robots loading items to be salvaged, and robots returning with or without items. 

In the remainder of this section,  each of the six SCSL language elements will be explained, using  the system test described in  Section~\ref{sec:systest} as an example.

While we give some intuitive explanations about the SCSL behavioural semantics in this section, a more formal comprehensive specification of semantics is presented in~\ref{sec:scslsem}.

\subsection{Primitive Types and Composite Data Types}

Each rover accepts commands from the command centre: initially, command $\gotodst$ will set a rover into motion, trying to reach the specified destination. Command $\pickup$ orders a rover that has reached the target destination to pick up an item to be salvaged. Command $\returntodst$ directs a rover to return to a specific position.\footnote{In principle, commands $\gotodst$ and $\returntodst$ have the same effect. We use two distinct commands, however, to clarify that first the target destination is specified to direct the robots to the goods to be salvaged (command $\gotodst$), and after that the destination is specified to let the robots return to a home base (command $\returntodst$).}

In the other communication direction, each rover sends status information to the command centre. While trying to reach the specified target destination, a rover is in state $\approaching$, but it may also get broken along the way (state $\fault$) or get stuck such that it will be unable to reach the destination (state $\stuck$). After a rover has successfully picked up an item at the target destination, it transits to state $\itemloaded$, and then to $\returningwi$ after having been commanded to return to a specified position. After arrival at the return position, the final state of rovers carrying items is $\loadedret$.
Some rovers will have to return without carrying items; for these, the states $\returning$ and $\returned$ apply.

These  enumeration types are declared as shown in Listing~\ref{lst:primtypes}.

\begin{lstlisting}[style=scsl,language=scsl,caption={Enumeration types.},label=lst:primtypes,mathescape=true]
enum
 RoverCommands : $\{ \gotodst, \returntodst, \pickup \}$;
 Status : $\{ \initial, \approaching, \stuck, \fault, \atdst, \itemloaded, $
            $\returning, \returningwi, \returned, \loadedret \}$;
end enum
\end{lstlisting}

We need a composite data type $\location$ for specifying geographic coordinates (e.g.~GPS) used for   target and return destinations sent to the rovers. Also, each rover can establish its own position in coordinates of type $\location$.
Another composite data type is $\area$ for specifying the exclusion zones that are present in the system testing area and should never be entered by any rover.\footnote{Typically, a $\area$ element would be represented as a simple polygon.}

\subsection{Global Constants}

The constants needed for our system test are specified in Listing~\ref{lst:const}.

\begin{lstlisting}[style=scsl,language=scsl,caption={Global constants.},label=lst:const,mathescape=true]
global const
  $n : \mathbb{N};$ -- initial number of rovers
  $m : \mathbb{N};$ -- number of items to be salvaged
  $k : \mathbb{N};$ -- minimal number of items to be salvaged
  constraint
    $k \le m \le n$
  end constraint
  $\tdst : \rnneg$; -- arrival deadline when at least $m$ rovers should 
               -- have arrived at destination
  $\tcend : \rnneg$; -- mission deadline when at least 
             -- $k$ items should have been salvaged
  constraint
    $\tdst \le \tcend$
  end constraint
  $\targetDst : \location$; -- target destination where the 
                         -- items to be salvaged can be found
  $\narea : \mathbb{N}$; -- number of exclusion zones in the testing range
  $\nga : \area[\narea]$; -- array of exclusion zones
  $\startPos : \location[n]$; -- start positions for each rover
  $\returnDst : \location[n]$; -- return destinations for each rover
  $\allIds : \mathbb \itemId^*$; -- List of item identifiers 
                    -- for each item to be salvaged
  constraint
    $\#\allIds = m$ -- there are $m$ items to be salvaged
  end constraint                  
end const
\end{lstlisting}

\subsection{Object Types}

The object types for our system test are rovers and command centres. Rovers accept commands $cmd$ from the control centre, the specification of destinations $dst$ to be reached by each rover, identifications $id$ of items to be picked up and salvaged, and their actual position $pos$ provided by some external positioning system like GPS. As output, each rover communicates its current status $s$ to the command centre. This interface specification is given in Listing~\ref{lst:rover}, lines~1 and~2. 

As motivated in Section~\ref{sec:intro}, original equipment, peripherals, and simulations thereof operate in fixed processing cycles that may differ between object types. Therefore a \emph{cycle time} is specified in relation to the (faster) observation and stimulation cycles that can be performed by the test equipment. The specification $\textsf{cycletime}\ 20$ states that any instance of type \textsl{Rover} performs one processing cycle in 20 observation cycles of the test equipment.

\begin{lstlisting}[style=scsl,language=scsl,caption={Object type `Rover'.},label=lst:rover,mathescape=true]
object type Rover($\inp cmd : \rcmds$,  $\inp dst : \location$, $\inp id : \itemId$,
                   $\inp pos : \location$, $\oup s : \status$)
   cycletime 20                 
end type
\end{lstlisting}

Command centres accept an array $s[i], i:0..(n-1)$, of status values, one from each rover, as inputs. They communicate possibly different commands $cmd[i]$, destination specifications $dst[i]$, and item identifications $id[i]$ to the rovers. This is specified in Listing~\ref{lst:cmdcentre}.

\begin{lstlisting}[style=scsl,language=scsl,caption={Object type `CommandCentre'.},label=lst:cmdcentre,mathescape=true]
object type CommandCentre( $\inp s : \status[n]$,
                             $\oup cmd : \rcmds[n]$,
                             $\oup dst : \location[n]$,
                             $\oup id : \itemId[n]$ )
   cycletime 10                          
end type      
\end{lstlisting}

%
%
%

Note that these object type specifications do not indicate how outputs of some objects will be linked to inputs of others,
since this cannot be determined once and for all on type level. Instead, the mapping from outputs to inputs is specified on object level in the collaboration part of the system test configuration (see Section~\ref{sec:systestconfig}).

\subsection{Auxiliary Object Types}\label{sec:auxobj}

While the object types introduced above represent constituents of the system under test, auxiliary object types can be introduced with the same syntax as introduced above for storing ``meta data'' related to the test execution status and performing auxiliary services to support the test execution. Instances of auxiliary object types are created in the testing environment and executed, for example, on original equipment harnesses or as parts of simulations.

\subsection{Global Function Declaration}

For evaluating the system test, a counting function is needed: $\numr(r,S)$ counts the number of rovers in array $r$ whose state has one of the values specified in set $S$. Boolean function $\innga(pos,\nga)$ returns $\ttt$ if and only if the geographic position $pos$ is in one of the exclusion zones $\nga[i],\ i:0..(\narea-1)$.\footnote{For determining whether a coordinate is inside a simple polygon, the \emph{ray casting algorithm} could be used~\cite{10.1145/356625.356626}.} Boolean function 
$\isclose(pos,dst)$ returns true if and only if a robot's actual position $pos$ is close to the destination coordinate $dst$, so that the robot is in the position to load an item.

\begin{lstlisting}[style=scsl,language=scsl,caption={Global functions.},label=lst:func,mathescape=true]
global function
  $\numr : \rover[n] \times \mathbb P(\status) \fun \mathbb N;$
  $(r,S) \mapsto \#\{ i : 0..(n-1)~|~r[i] \neq \scnull \wedge r[i].s \in S   \};$
  $\innga : \location \times \area[\narea] \fun \mathbb B;$
  -- $\innga(pos,\nga) = \ttt$ iff $pos$ is contained in one 
  -- of the exclusion zones $\nga[i]$ 
  -- (application of the ray casting algorithm)
  $\isclose : \location \times \location \fun \mathbb B;$
  -- $\isclose(pos,dst) = \ttt$ iff $pos$ is close to $dst$
end function
\end{lstlisting}

For functions $\innga$ and $\isclose$, the detailed specifications have been ommitted in Listing~\ref{lst:func}. They would also be specified in Z-style, as illustrated in the case of the function $\numr$.

\subsection{Elementary Scenario Types}

\paragraph{Scenario structure and behavioural specification}
Elementary scenario types specify how the objects of interest should behave or interact with each other in certain operational situations. In SCSL, each elementary scenario type comes with the following sub-components.
\begin{enumerate}
\item An interface specification containing parameters representing object references\footnote{Similar to Java, instances of object types are passed by reference to the elementary SCSL scenarios} and scalar input values used for scenario configuration. 

\item A precondition over interface parameters, specifying the conditions of an elementary scenario instance to start its execution.
\item A behavioural specification consisting of
\begin{enumerate}
\item LTL formulae over interface parameters and auxiliary variables,
\item optional initial actions setting the initial values of auxiliary variables,
\item optional condition-action pairs defining writes to auxiliary variables, depending on conditions over interface parameters and auxiliary variables.
\end{enumerate}
\end{enumerate}

For elementary scenario types, all parameters are inputs, so keyword ``$\mathbf{in}$'' is superfluous: Auxiliary outputs to be relayed to other scenarios at runtime are communicated via auxiliary objects (Section~\ref{sec:auxobj}).
Conditions in condition-action pairs  can be   
\begin{itemize}
\item \emph{guards} $g$ that trigger the associated action in every execution cycle where $g$ evaluates to $\ttt$. 
\item \emph{change conditions} $chg$ that trigger the associated action in every execution cycle where $chg$ changes from $\fff$ to $\ttt$.
\end{itemize}

This concept for behavioural modelling deliberately mixes declarative and imperative specification styles: LTL formulae are implicit (=declarative) behavioural specifications that are refined by adding conditions and actions in  imperative 
style that is quite similar to state machines used in UML~\cite{uml_2_5}. From our experience, LTL formulae (or declarative specifications in general) often specify behaviours that are too general, so that some unwanted executions are still accepted by these formulae. The condition-action pairs changing auxiliary variables and the use of the latter in LTL formulae allows to narrow down the accepted behaviours in an effective way.

The exclusion of unwanted behaviours is additionally supported by the utilisation of \emph{frames}, as originally introduced in the Vienna Design Method VDM~\cite{jonesVDM}. Pre-defined set-valued auxiliary variable $\frm$  contains the names of all object parameters that may be changed in the current execution cycle. This helps in situations where the SUT should conform to a declarative specification $\varphi(o_1.v_1, o_2.v_2,\dots,o_k.v_k)$ without changing a subset of the object parameters $o_i.v_j$. The symbol names of changeable object parameters must  always be contained in variable $\frm$. Auxiliary variable $\frm$ is implicitly set to the complete collection of all parameters of all objects referenced in the scenario parameter list. It can be changed in the actions specified for the scenario type by adding or removing object parameter symbols.

Moreover, each elementary scenario is associated with an implicitly defined variable $\sactive$ that is set to $\ttt$ when the precondition evaluates to $\ttt$ for the first time. When  $\sactive$ is reset to $\fff$, this indicates the termination of the scenario execution. After scenario termination, the behavioural specifications need to hold no longer. One global variable $\tend$ is $\fff$ at the beginning of a system test execution, but can be set to $\ttt$ by any elementary scenario, whereupon the system test terminates. If $\tend$ is never set to $\ttt$, a system test ends when all elementary scenarios involved have terminated (status change from $\sactive = \ttt$ to $\sactive = \fff$).

\paragraph{Rover-specific scenario types}
For the system test under consideration, three elementary scenario types cover the rover-specific behaviours: Scenario type
\textsl{Approach} specifies rover behaviour while approaching the target destination. Scenario
\textsl{Pickup} specifies rover behaviour while picking up an item to be salvaged. Scenario  \textsl{Return} specifies rover behaviour while returning to a specified location, with or without a loaded item.

The \textsl{Approach} scenario type is declared in Listing~\ref{lst:approach}. As input, each \textsl{Approach}-instance gets a reference to a specific rover $r$ and the location $\startPos$ from where the rover shall start its journey to the target destination. The latter is specified in $r.dst$, and this destination is communicated  from the command centre to the rover,
before the scenario becomes operative. 

The precondition for an \textsl{Approach} scenario instance to become active states that (1)~the rover's initial state is $\initial$ (this is an expected behaviour when the rover is started), (2)~the rover must have been placed at its starting position (this must be ensured by the system testing crew), and (3)~the command input to the rover must be $\gotodst$ (so a scenario instance can only become active after the command centre has sent this command to the rover).

For scenario type \textsl{Approach}, the   behaviour is specified by LTL formulae only -- no condition-action specifications are necessary: the first formula specifies the scenario termination condition. It states that $\sactive$ will be set to $\fff$ after either (1)~the rover inadvertently enters an exclusion zone (this actually destroys the rover, as will be handled in scenario type \textsl{MishapHandler}), or (2)~the rover actually reaches the specified destination or gets stuck or broken before reaching it, or (3)~the command centre sends a new command to the rover, indicating that reaching the target destination is no longer necessary.
The second formula asserts that the rover will finally get close to the target destination $r.dst$ and change its state to $r.s = \atdst$, unless it vanishes in an exclusion zone, gets broken or stuck, or receives another directive from the command centre.

Scenario type \textsl{Approach} is an \emph{observer scenario}. It specifies expected behaviour only, and instances of \textsl{Approach} must not change any object parameters. Therefore, the empty set is assigned to the $\frm$-variable in the initial action.  The original equipment harness running the scenario instance of \textsl{Approach} for a specific rover would signal an error if the second specification does not become $\ttt$ before the system test terminates.

\begin{lstlisting}[style=scsl,language=scsl,caption={Approach scenario.},label=lst:approach,mathescape=true]
elementary scenario Approach( $r$ : Rover, $\const \startPos$ : Location )
  precondition $r.s = \initial \wedge \isclose(r.pos,\startPos) \wedge r.cmd = \gotodst$;
  spec $\tg \big( (\innga(r.pos,\nga)  \vee  r.s \in \{ \stuck, \fault, \atdst \} \vee {}$
                 $r.cmd \neq \gotodst ) \Rightarrow \tx\neg \sactive\big)$;
  spec $\tf \big(  \innga(r.pos,\nga)  \vee  r.s \in \{ \stuck, \fault \} \vee {}$
                 $r.cmd \neq \gotodst \vee (\isclose(r.pos,r.dst) \wedge r.s = \atdst) \big)$;   
  initact $\frm := \varnothing;$                            
end scenario      
\end{lstlisting}

Scenario type \textsl{Pickup} (Listing~\ref{lst:pickup}) specifies the expected rover behaviour during the pickup phase. Its precondition to become active is that the rover is commanded to either pick up an item or return to a given return destination. In the latter case, the first LTL formula states that the scenario will immediately become inactive, since there is no item to be picked up. Other scenario termination conditions are that the rover enters an exclusion zone or gets stuck or broken while trying to load the item, or that the item has been successfully loaded. If the command $\pickup$ applies, the rover will finally succeed in loading the item, unless it fails before. This is expressed by the second LTL formula. Again,    \textsl{Pickup} is an  observer scenario type.

\begin{lstlisting}[style=scsl,language=scsl,caption={Pickup scenario.},label=lst:pickup,mathescape=true]
elementary scenario Pickup( $r$ : Rover )
  precondition $r.cmd \in \{  \pickup, \returntodst \} $;
  spec $\tg \big( (r.cmd = \returntodst \vee \innga(r.pos,\nga) \vee  {}$
                    $ r.s \in \{ \stuck, \fault, \itemloaded \}) \Rightarrow   \tx\neg \sactive\big)$;
  spec $r.cmd = \pickup \Rightarrow  {}$
       $\tf \big( \innga(r.pos,\nga) \vee r.s \in \{ \stuck, \fault, \itemloaded \} \big);$  
  initact $\frm := \varnothing;$                                    
end scenario      
\end{lstlisting}

Observer scenario type \textsl{Return} (Listing~\ref{lst:return}) specifies how a rover returns to a given address. 
It becomes active as soon as a $\returntodst$ command has been received from the command centre.
Here we have an initial action that introduces a Boolean auxiliary variable $aux_\text{isLoaded}$ (its type is determined automatically from the right-hand side part of the assignment expression) which is set to $\ttt$ if and only if
the rover has loaded an item.
As usual, the rover may be lost after entering an exclusion zone, or it may get stuck or broken along the way. If, however, all goes well, it will finally arrive at the return destination and assume status $\loadedret$ or $\returned$, depending on whether it returns with an item or not. While approaching the return destination, the rover state $r.s$  is
$\returningwi$ or $\returning$, depending on the value of $aux_\text{isLoaded}$.

Note that auxiliary variables need not be declared: their type is specified implicitly from the right-hand-side expressions in the assignments, similar to {\tt auto} declarations in \cpp.

\begin{lstlisting}[style=scsl,language=scsl,caption={Return scenario.},label=lst:return,mathescape=true]
elementary scenario Return( $r$ : Rover )
  precondition $r.cmd = \returntodst$;
  spec $\tg \big( (\innga(r.pos,\nga) \vee {}$
                $r.s \in \{ \stuck, \fault, \returned,\loadedret \})   $
                                        $ {} \Rightarrow \tx\neg \sactive\big)$;
  spec $\big( (r.s = \returningwi \wedge aux_\text{isLoaded} ) \vee (r.s = \returning \wedge \neg aux_\text{isLoaded} )  \big)$
        $\tu$
        $\big( \innga(r.pos,\nga) \vee r.s \in \{ \stuck, \fault \} \vee {}$
          $(\isclose(r.pos,r.dst) \wedge aux_\text{isLoaded} \wedge r.s = \loadedret   ) \vee {}$
          $(\isclose(r.pos,r.dst) \wedge \neg aux_\text{isLoaded} \wedge r.s = \returned) \big)$;                                 
  initact $\frm := \varnothing;\ aux_\text{isLoaded} := (r.s  =  \itemloaded)$;    
end scenario      
\end{lstlisting}

\paragraph{Scenarios for interactions between rovers and command centre}

We need four elementary scenario types for describing the interaction between the command centre and the rovers during their approach to the items to be salvaged (\textsl{ApproachHandler}), while picking up items (\textsl{PickupHandler}), while returning to their specified destinations (\textsl{ReturnHandler}), and in the special case where a rover inadvertently enters an exclusion zone (\textsl{MishapHandler}).

\newpage

\begin{lstlisting}[style=scsl,language=scsl,caption={Scenario  for rovers approaching the target destination.},label=lst:scapproach,mathescape=true]
elementary scenario ApproachHandler( $cc$ : CommandCentre,
                                       $r$ : Rover,
                                       $i$ : $\mathbb N_0$ )
  spec $\tg \big( \innga(r[i].pos,\nga) \Rightarrow \tx\neg \sactive\big)$;
  spec  $\tg \big( r[i].s \in \{ \initial, \approaching \} \Rightarrow  {}$
                         $(cc.cmd[i] = \gotodst \wedge cc.dst[i]  =  \targetDst)  \big)$;    
  initact $\frm := \{ cc.cmd[i], cc.dst[i] \};$                       
end scenario      
\end{lstlisting}

Scenario type \textsl{ApproachHandler} (Listing~\ref{lst:scapproach}) inputs references to the command centre, the robots, and an index of the specific rover $r[i]$ whose interaction with the command centre is considered in an \textsl{ApproachHandler} instance. The first LTL formula specifies that a scenario instance will terminate immediately if rover $r[i]$ enters an exclusion zone, since the rover will be lost and can never reach the target destination. The second LTL formula asserts that when the rover has just been initialised or while it  is still approaching the target destination, the command centre will continuously transmit the $\gotodst$ command and the $\targetDst$ (location of the items to be salvaged) to the rover. Note again that the transmission interfaces from $cc.cmd[i]$ to $r[i].cmd$ and from $cc.dst[i]$ to $r[i].dst$ are not specified in the elementary scenario declarations,  but will be specified for the concrete objects in the collaboration which is part of the test configuration shown below.

Elementary scenario type \textsl{PickupHandler} (Listing~\ref{lst:scpickup}) specifies how the command centre directs rovers having arrived at the target destination to load items to be salvaged. To this end, three auxiliary variables are initialised (line~8):
\begin{enumerate}
\item $aux_\text{id}$ is initialised with the list $\allIds$ containing identifications of all items to be salvaged.
\item $aux_\text{rAtDst}$ is initialised as the empty set and then used to store all indexes of rovers that have arrived at the target destination, but   have not yet been assigned an item to carry (line~9).
\item $aux_\text{rLoading}$  is initialised as the empty set and then used to store all indexes of rovers that have been assigned to carry an item but have not yet completed the task of loading the item (lines~10---11).
\end{enumerate}

%

\newpage
\begin{lstlisting}[style=scsl,language=scsl,caption={Scenario  for rovers commanded to pick up an item.},label=lst:scpickup,mathescape=true]
elementary scenario PickupHandler( $cc$ : CommandCentre, 
                                     $r$ : Rover[n] )
  spec  $\bigwedge_{i:0..(n-1)} \tg\Big( r[i] \neq \scnull \Rightarrow \big(  i\in aux_\text{rLoading} \Rightarrow {}$
                         $ (cc.cmd[i] = \pickup \wedge cc.id[i] = aux_\text{loadItemId}[i] ) \big) \Big)$;  
  spec $\tg\big( aux_\text{id} = \varepsilon \vee {} $
          $(aux_\text{rAtDst} = \varnothing \wedge \numr(r,\{\initial, \approaching \}) = 0) \Rightarrow {}$
                       $\tx \neg \sactive \big)$;                         
  initact $\frm := \varnothing;\ aux_\text{id} := \allIds$; $aux_\text{rAtDst} := \varnothing$; $aux_\text{rLoading} := \varnothing$;
  cndact $[ \ttt ]$ / $aux_\text{rAtDst} := \{ i:0..(n-1)~|~r[i].s = \atdst \} \setminus aux_\text{rLoading}$;
    $aux_\text{rLoading} := aux_\text{rLoading} \setminus \{ i:0..(n-1)~|~r[i].s \in\{ \itemloaded, \stuck, \fault\} \}$;
    if $( aux_\text{id} \neq\varepsilon \wedge aux_\text{rAtDst} \neq\varnothing )$  then 
     $\ell := \min aux_\text{rAtDst}$; 
     $aux_\text{rAtDst} := aux_\text{rAtDst} \setminus \{\ell\}$;
     $aux_\text{rLoading} := aux_\text{rLoading}\cup \{ \ell \}$;
     $aux_\text{loadItemId}[\ell] := \text{\tt popfront}(aux_\text{id})$;
   endif
   $\frm := \{ cc.cmd[i],cc.id[i]~|~i \in aux_\text{rLoading} \}$;
end scenario      
\end{lstlisting}

The assignment of items to rovers is performed in the condition-action expression in lines~11---16: as long as there are still items to be loaded ($aux_\text{id} \neq\varepsilon$), and whenever at least one rover is located at the target destination but still without an assigned item ($aux_\text{rAtDst} \neq\varnothing$), the following action is performed:
\begin{enumerate}
\item The rover with the smallest index $\ell$ in $aux_\text{rAtDst}$ is selected to load the next item in the list.
\item This index $\ell$ is removed from the set $aux_\text{rAtDst}$ of rovers that could still load an item and added to the 
set $aux_\text{rLoading}$  of rovers currently loading an item.
\item In auxiliary array $aux_\text{loadItemId}$, the id of the next item to be loaded is written to index $\ell$ and removed from the list of items still to be salvaged (this is performed by instruction $\text{\tt popfront}(aux_\text{id})$ which returns the list head and removes it from the list at the same time).
\end{enumerate}

The first LTL formula (lines~3---4) specifies that the command centre sends a $\pickup$ command to all rovers that still exist 
in the collaboration ($r[i] \neq \scnull$) and whose index has been inserted into set $aux_\text{rLoading}$. Moreover, the command centre provides the associated item id in output parameter $cc.id[i]$.
The second LTL formula specifies that the  scenario shall terminate after all items have been assigned to rovers for loading 
or if there aren't any rovers left to carry an item. Since scenario \textsl{PickupHandler} acts as a simulation, the $\frm$-variable must be set in every cycle in such a way that the scenario is allowed to set $cc.cmd[i]$ and $cc.id[i]$ can be set by this simulations. Therefore, the frame variable is always set in line~17 to   
all $cc.cmd[i], cc.id[i]$ where rover $i$ is contained in set $aux_\text{rLoading}$.

Elementary scenario type \textsl{ReturnHandler}  (Listing~\ref{lst:cmdfault}) specifies the return conditions and the associated commands from the command centre to the rovers. Again, this is a simulation scenario, as explained in Section~\ref{sec:systest}.    Return commands will be finally sent to all rovers that are still part of the (dynamically changing) system test configuration. This is checked by condition $r[i] \neq \scnull$: a rover that is no longer existent will be marked by $r[i] = \scnull$ in the collaboration part of the configuration. The LTL formula in lines~3 and~4 specifies that return commands will be immediately sent to any rover that has loaded its designated item to be salvaged (it is then in state $\itemloaded$) or that is stuck or broken. In the latter cases it is unclear whether the robot will still be able to execute the return command, but the command is at least issued by the centre. A return command $\returntodst$ to rover $r[i]$ is associated with a destination value $\returnDst[i]$ which is a predefined constant.

The second LTL formula in lines~5 to~10 specifies return commands to rovers that are still operative, approach the target destination or have already arrived there. These rovers are commanded to return if they are no longer needed, because all items to be salvaged are already being transported by other rovers.

\begin{lstlisting}[style=scsl,language=scsl,caption={Handler for   rovers that should return to a specified destination.},label=lst:cmdfault,mathescape=true]
elementary scenario ReturnHandler( $cc$ : CommandCentre,
                                     $r$ : Rover$[n]$ )
  spec  $\bigwedge_{i:0..(n-1)}\tg \big( r[i] \neq \scnull \Rightarrow (r[i].s \in \{ \stuck, \fault, \itemloaded \} \Rightarrow  {}$
                         $\tx (cc.cmd[i] = \returntodst \wedge cc.dst[i]  =  \returnDst[i]))  \big)$;    
  spec  $\bigwedge_{i:0..(n-1)}\tg \big( r[i] \neq \scnull \Rightarrow {}$
        $(r[i].s \in \{ \approaching, \atdst \} \wedge {}$
          $\numr(r,\{\itemloaded \}) + {}$
          $\numr(r,\{ \returningwi \}) + {}$
          $\numr(r,\{ \loadedret \}) = m \Rightarrow  {}$
                         $\tx (cc.cmd[i] = \returntodst \wedge cc.dst[i]  =  \returnDst[i]))  \big)$;  
  initact $\frm := \{ cc.cmd[i],cc.id[i]~|~i = 0,\dots,(n-1)\};\ $                                              
end scenario      
\end{lstlisting}

The effect of rovers entering exclusion zones is specified by scenario type \textsl{MishapHandler}. Such robots have to be removed from the system test collaboration, since they are no longer existent. This is expressed by the change condition and action in line~3 of the scenario: when a robot enters an exclusion zone for the first time, action $\mathsf{delete}(r)$   is performed on the collaboration $coll$. This resets the reference $r$ to the rover to $\scnull$ and deletes all interfaces connecting  $r$-parameters to parameters of any other object. After this, the scenario is terminated, as specified by the LTL formula in line~2.

\begin{lstlisting}[style=scsl,language=scsl,caption={Mishap handler scenario.},label=lst:mishaps,mathescape=true]
elementary scenario MishapHandler( $r$ : Rover, $coll$ : collaboration )
  spec $\tg \big( \innga(r.pos,\nga) \Rightarrow \tx\neg \sactive\big)$;
  cndact $\chgc{\innga(r.pos,\nga)} / coll.\mathsf{delete}(r)$;     
end scenario      
\end{lstlisting}

\paragraph{Scenario for checking emergent properties}
The evaluation of pass criteria for a system test is also specified in one or more elementary scenario types. For the test discussed here, the pass criteria are specified in the \textsl{EmergentPropertyChecker} scenario type in Listing~\ref{lst:emergentchk}. For a test to pass,
\begin{enumerate}
\item at least $m$ rovers shall arrive at the salvage destination before time $\tdst$, and
\item at least $k$ rovers shall return with a salvaged item before time $\tcend$.
\end{enumerate}
The system test is terminated at time $\tcend + 10$. The built-in variable $\tsc$ denotes the current test execution time, starting with zero when the test execution begins.

\begin{lstlisting}[style=scsl,language=scsl,caption={Emergent property checker.},label=lst:emergentchk,mathescape=true]
elementary scenario EmergentPropertyChecker( $r$ : Rover[n] )
  spec  $(\tsc < \tdst)~\tu~(\numr(r,\{\atdst,\itemloaded,\returning,$
                        $\returningwi,\returned,\loadedret \}) \ge m)$;
  spec  $(\tsc < \tcend)~\tu~(\numr(r,\{ \loadedret \}) \ge k)$;
  spec  $ \tg\big( \tsc = \tcend + 10  \Rightarrow \tx \tend \big)  $;
  initact $\frm := \varnothing;$
end scenario      
\end{lstlisting}

\paragraph{Remark -- specification of object behaviour}
Note that in principle, preconditions and behavioural specifications can also be defined for object types -- this is syntactically and semantically well-defined. However, object type behaviours should only be defined if they hold in \textit{every} elementary scenario instance an object is involved in. For the system test example discussed here, we could not identify any behaviours that would hold in arbitrary operational situations. We suspect that this is very likely to hold in most scenario-based test specification for cyber-physical systems.

%

\subsection{System Test Configuration and Deployment} \label{sec:systestconfig}

A system test configuration represents a \emph{composite scenario}: it specifies how SUT objects and instances of elementary scenarios should interact to perform a system test. 
For the system test described in Section~\ref{sec:systest}, the configuration is shown in Listing~\ref{lst:systest}.

\begin{lstlisting}[style=scsl,language=scsl,caption={System test configuration.},label=lst:systest,mathescape=true]
systemtest
  coll : collaboration
    $r$ : Rover[n];
    $cc$ : CommandCentre;
  
    interface $\mathit{Is}[i]$ from $r[i].s$ to $cc.s[i]$ for $i:0..(n-1)$;
    interface $\mathit{Icmd}[i]$ from $cc.cmd[i]$ to $r[i].cmd$ for $i:0..(n-1)$;
    interface $\mathit{Idst}[i]$  from $cc.dst[i]$ to $r[i].dst$ for $i:0..(n-1)$;
    interface $\mathit{Iid}[i]$  from $cc.id[i]$ to $r[i].id$ for $i:0..(n-1)$;
  end collaboration      

  schedule
    $\parallel_{i:0..(n-1)}$ $\Big($Approach(coll.$r[i]$,$\startPos[i]$);Pickup(coll.$r[i]$);Return(coll.$r[i]$);$\Big)$
    $\parallel_{i:0..(n-1)}$  ApproachHandler(coll.$cc$,coll.$r[i]$,$i$)
    $\parallel_{i:0..(n-1)}$  MishapHandler(coll.$r[i]$,coll)
    $\parallel$ PickupHandler(coll.$cc$,coll.$r$)
    $\parallel$ ReturnHandler(coll.$cc$,coll.$r$)
    $\parallel$ EmergentPropertyChecker(coll.$r$)
  end schedule
end systemtest      
\end{lstlisting}

\paragraph{The collaboration}

The collaboration part of our system test specification introduces $n$ robots $r[i], i:0..(n-1)$ of object type \textsl{Rover} and one command centre $cc$ of type \textsl{CommandCentre}. Object types are similar to classes in programming languages.  

The interfaces declared in a collaboration specify how output parameters of certain objects are mapped to input parameters of others. For example,   interface $\mathit{Is}[i]$ specifies that the output parameter $s$ (for ``status'') of rover $r[i]$ is mapped to input parameter $s[i]$  of the command centre~$cc$ ($cc.s$ is a status array of length $n$). Since all objects collaborating in a system test operate in a cyclic mode, the semantics of the interface mappings is fairly easy:
\begin{itemize}
\item A new parameter value written to $s$ by $r[i]$ in cycle $p$ becomes visible at $cc.s[i]$ at the start of $r[i]$'s execution cycle $p+1$.
\item If $cc$ operates with a slower cycle time than $r[i]$, then the most recent value written by $r[i]$ to $s$ becomes visible at the start of $cc$'s next cycle, so previous writes may be lost.
\item If $cc$ cycles faster than $r[i]$, it may read the same value several times (interfaces are like shared variables).
\item If two objects communicate over a communication medium that needs more time than a single cycle to deliver new values to their destination, an auxiliary object with a slower cycle time has to be introduced. This object represents the communication medium.
\end{itemize}

An essential characteristic of system tests for interacting autonomous robots\footnote{or, for \emph{systems of systems} in general~\cite{DBLP:journals/csur/NielsenLFWP15}} is that the collaboration may change during the test execution. In the example discussed here, robots can get lost when inadvertently entering an exclusion zone. In other system tests, new robots might enter the collaboration or leave it in an orderly fashion. Therefore, collaborations are semantically represented as abstract data types allowing operations like
\begin{enumerate}
\item $coll.\mathsf{delete}(r[k])$. Delete rover  $r[k]$ from the collaboration, including all interfaces $r[k]$ contributes to. The deletion is marked by setting $coll.r[k] = \scnull$. This helps to check in elementary scenarios whether a robot still exists in the collaboration.
\item $coll.\mathsf{create}(r[n+\ell] : \text{Rover})$. Extend the collaboration by  a new rover object that is registered
under the new array index $n+\ell$.\footnote{Arrays in SCSL are variable-size containers whose elements can be accessed via indexes, similar to \texttt{ArrayList} objects in Java.}
\item $coll.\mathsf{create}(\mathsf{interface}\ \mathit{Is}[n+\ell]\ \mathsf{from}\ r[n+\ell].s\  \mathsf{to}\ cc.s[n+\ell]  )$
creates a new interface in the collaboration.

\end{enumerate}
These operations are atomic, so the collaboration abstract data type acts like a monitor. This is necessary since several concurrent elementary scenarios may request configuration changes simultaneously during their execution.

\paragraph{Remark -- auxiliary objects}
The robotic system test example discussed here does not require auxiliary object types (Section~\ref{sec:auxobj}) for communicating test execution data between elementary scenarios. If needed, instances of auxiliary object types and interfaces between such objects are declared just as SUT-related objects in the  collaboration section.

\paragraph{The schedule}

The schedule part of a system test configuration specifies the test execution by means of elementary scenario instances that are sequentially composed or run in parallel. For example, \textsl{Approach}($coll.r[i]$,$\startPos[i]$) specifies an instance of elementary scenario type \textsl{Approach}(\dots), where the formal parameter $r$ has been substituted by the concrete 
rover instance $coll.r[i]$ that is part of the collaboration. Concrete parameter $\startPos[i]$ is an element of a constant array specifying the starting positions from where each rover commences its salvage expedition. After \textsl{Approach}($coll.r[i]$,$\startPos[i]$) has terminated, scenario instance \textsl{Pickup}($coll.r[i]$) can execute, after which 
instance \textsl{Return}($coll.r[i]$) runs. This sequential execution of three elementary scenarios is performed concurrently for each rover $coll.r[i],\ i:0..(n-1)$.
In a similar way, instances of the remaining elementary scenarios are specified to be executed concurrently.

\emph{System test deployment} denotes the task of mapping the logical components of a system test -- these are identified in the system test configuration -- to software procedures, threads, processes, hardware interfaces, and processors. The details of system test deployment are beyond the scope of this paper, but we indicate some important concepts and variants.

\begin{figure}[H]
\begin{center}
\includegraphics[width=\textwidth]{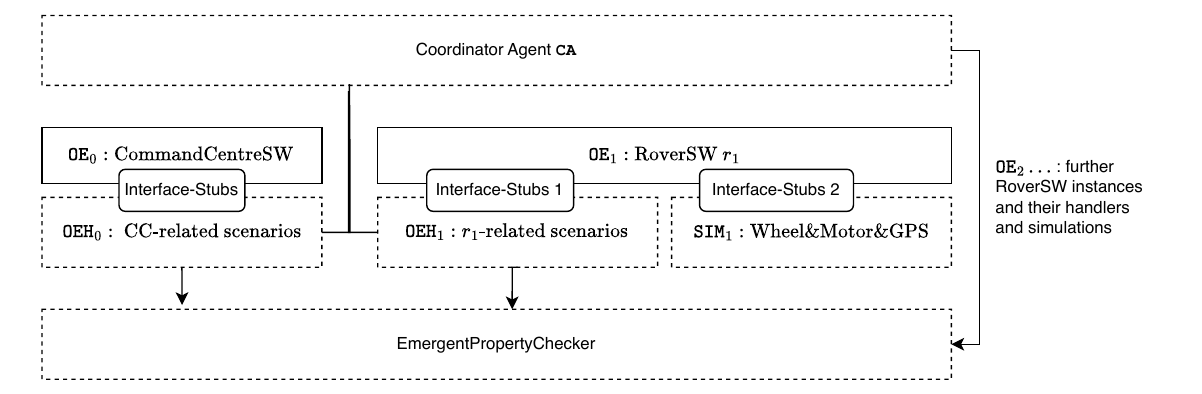}
\end{center}
\caption{Instantiation of Fig.~\ref{fig:systestconfig} for the `software-only' system test deployment.}
\label{fig:swonlyconfig}
\end{figure}

\paragraph{Deployment variant  `software only'}

System tests like the one described in this paper are usually quite costly to  be performed in the real world. Therefore, tests are usually performed in several phases, each phase using a different deployment. The first test phase is typically  ``software  only'', where no target hardware is involved. 
In this setting, the original equipment $\OEQ_0, \OEQ_1, \dots$   consists only of the software of the command centre and the robots, as sketched in Fig.~\ref{fig:swonlyconfig} which specialises the generic test configuration diagram of  Fig.~\ref{fig:systestconfig} for the software-only deployment of our system test. 

The communication interfaces of each robot are stubbed, since the hardware interfaces and peripherals are not available in the software-only configuration. This allows to realise the interface declarations shown in the system test configuration by shared variables, procedure calls, socket communication, or similar means of data exchange between software components. Consequently, any deployment requires a specification explaining how the interfaces identified in the system test configuration should be implemented. 
Also, since the \OEQ software does not run on original hardware, a deployment configuration needs to map the $\OEQ_i$ to processors available for the software test execution.\footnote{For example, the software-only tests could be executed in the cloud, and each $\OEQ_i$ could be encapsulated  in a separate docker container that is executed in a Kubernetes POD of its own, see \url{https://kubernetes.io/docs/concepts/workloads/pods/}.}

Since the hardware peripherals are missing in the software-only deployment, their feedback to the \OEQ needs to be simulated. For example, each rover's outputs to motors setting wheel angles and wheel speed is caught by a software simulation (see ${\tt SIM}_1$ in Fig.~\ref{fig:swonlyconfig}) that calculates the rover's position changes that would result from these motor actuations in the real world. The calculated positions would be passed to the rover software, simulating satellite positioning information that would be received in the real world.

The original equipment harnesses $\OEH_0, \OEH_1 \dots$ each execute a subset of the elementary scenario instances identified
in the system test schedule. For instance, $\OEH_0$, the handler for the command centre, would execute
\begin{equation*}
\begin{array}{l}
     \parallel_{i:0..(n-1)}   \text{\sl ApproachHandler(coll.cc,coll.r[i],i)} \\
     \parallel  \text{\sl PickupHandler(coll.cc,coll.r)} \\
    \parallel  \text{\sl ReturnHandler(coll.cc,coll.r)},
    \end{array}
\end{equation*}
while the rover-specific $\OEH_i,\ i = 1,2, \dots$ would execute
\begin{equation*}
\begin{array}{l}
     \Big(\text{\sl Approach(coll.r[i],\startPos[i]);Pickup(coll.r[i]);Return(coll.r[i]);}\Big)
     \\
     \parallel \text{\sl MishapHandler(coll.r[i],coll)}.
    \end{array}
\end{equation*}
A global scenario execution unit executes the $\text{\sl EmergentPropertyChecker(coll.r)}$ instance.
The coordinator agent monitors requested changes of the collaboration and provides the actual collaboration state to all  elementary scenario instances under execution.



\paragraph{Deployment variant `real-world'}

For testing  the salvage mission described above in the real world, the \OEQ instances consist of the command centre, the complete robots (hardware with embedded software), the real communication service (for example,  satellite-based or radio-based message exchange), and the real location service (e.g.~GPS or Galileo). The only simulation left in a fully automated real-world system test would  mimic users in the  command centre initiating the salvage mission,  as all physical hardware, sensors, and communication channels are real.


\section{The SCSL Tool Platform}\label{sec:tools}

The SCSL Tool Platform consists of two main components. The first component is responsible for the generation of test artifacts from SCSL specifications, including test suites, runtime monitors, executable oracles, and simulations. A single specification language is used to derive these different artifacts, enabling a unified and consistent development process. In particular, LTL-based runtime monitors and executable oracles are used to assess the correctness of the SUT during execution. Oracles are executed in parallel within test executions to verify the observed behavior of the SUT, while simulations actively stimulate the SUT to trigger additional state transitions and expose potential failure behavior.

The second component of the SCSL Tool Platform is a cloud-based execution environment that orchestrates simulations, oracles, and test executions against the SUT. This environment is based on a time-synchronised, UDP-based, eventually consistent shared-state communication model for distributed runtime verification. It is composed of multiple agents, each deployed as a Docker container\footnote{Docker is a virtualisation engine that enables applications to run in lightweight, portable containers with all their dependencies isolated from the host system; for reference visit \url{https://www.docker.com/} }. Depending on its role, an agent may execute a simulation, an oracle, a test executor, or provide a virtualised environment for running the SUT software.

Inter-agent communication follows a blackboard-style architecture implemented as a time-triggered shared communication state. Each agent continuously publishes its output state as JSON\footnote{JSON (JavaScript Object Notation) is a lightweight, text-based data exchange format based on key–value pairs; for reference, see \url{https://www.json.org/}}-encoded messages via UDP multicast over a Docker MACVLAN network\footnote{MACVLAN is a special Docker network driver that assigns each container a unique MAC address, allowing it to appear as a device on the physical local network without significant virtualization overhead, thereby enabling fast, near real-time communication; for reference, see \url{https://docs.docker.com/engine/network/drivers/macvlan/}}. Incoming messages are received asynchronously and aggregated into a local input state. At each global time tick, agents create a snapshot of this input state, which is then used for deterministic local evaluation.

Both components are described in more detail in the following sections.

\subsection{SCSL-Based Test Artifact Generation}

The SCSL Artifact Generator follows a three-layer architectural paradigm. In the first layer, a parser processes the scenario specification files, including scenario definitions, scheduling rules, and variable/class type declarations. As output, the parser produces a structured intermediate representation consisting of 
\begin{itemize}
	\item a set of elementary scenario specifications,
	\item a set of instance specifications, 
	\item a directed acyclic graph of scenario scheduling rules and
	\item a set of typed variable definitions.
\end{itemize}

\begin{figure}[H]
	\includegraphics[width=\textwidth]{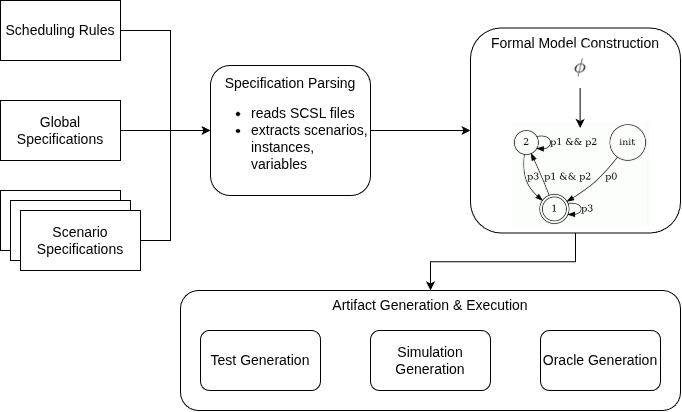}
	\centering
	\caption{SCSL-Based Test Artifact Generation Workflow}
	\label{fig:artifactgenwf}
\end{figure}

In the second layer, each parsed scenario specification is transformed into a formal behavioral representation. For each scenario, the associated precondition and behavioral constraints are combined into a temporal logic specification. In particular, whenever the defined precondition is satisfied, all associated behavioral specifications are required to hold in the subsequent time step. This relationship is formalised as:

$$
\phi_{\text{scsl}} \equiv \texttt{precondition} \;\wedge\; \tx\left(\bigwedge_i \texttt{specification}_i\right)
$$

Note that at this stage, only the scenario specification and the precondition are considered; condition/action items and scenario parameters are not yet incorporated. Condition/action items are evaluated later during test generation, simulation execution, or oracle evaluation, while scenario parameters are handled in a subsequent step.

The resulting formula $\phi_{\text{scsl}}$ is then abstracted into a propositional form $\phi_{\text{scsl-abstracted}}$, where complex expressions and function calls are replaced by atomic Boolean propositions, as in the example shown below.

$$
\begin{array}{rcl}
	\phi_{\text{scsl}}
	& = &
	(x > 0)\;\wedge\; \mathbf{X}\Big((x = z)\;\mathbf{U}\;(x < 100)\Big) \\[6pt]
	
	\Rightarrow \quad
	\phi_{\text{scsl-abstracted}}
	& = &
	p_0 \;\wedge\; \mathbf{X}\big(p_1 \;\mathbf{U}\; p_2\big)
\end{array}
$$

This step of propositional abstraction mapping enables automated synthesis using standard linear temporal logic (LTL) tools. Subsequently, the well-known tool \texttt{ltl3ba}~\cite{10.1007/978-3-319-11737-9_24} is used to translate the abstracted LTL formula into a \buchi automaton.

The resulting automaton is then analysed by the SCSL Artifact Generator to construct an internal symbolic \buchi automaton. More precisely, \texttt{ltl3ba} outputs a Hanoi $\omega$-Automaton (HOA) representation of the \buchi automaton corresponding to $\phi_{\text{scsl-abstracted}}$. This representation is parsed using the HOAF parser library\footnote{\url{https://automata.tools/hoa/cpphoafparser/}}. During this step, the atomic propositions associated with each transition are mapped back to their corresponding concrete expressions, yielding a symbolic \buchi automaton with transition guards represented as quantifier-free first-order expressions.

Based on the instance specifications defined in the scheduling rules, concrete scenario instances are created by instantiating elementary scenarios with unique instance identifiers and parameter assignments. Each instance therefore represents a parameterised realisation of a scenario template.

To ensure correct usage of scenario parameters during execution, parameter assignments are systematically integrated into the behavioral model. In particular, the conjunction of all parameter variable assignments associated with an instance is incorporated into each transition guard of the corresponding symbolic \buchi automaton. As a result, transitions are only enabled when both the original behavioral conditions and the instance-specific parameter constraints are satisfied. This guarantees that the behavior of each instantiated scenario is evaluated consistently with respect to its parameterisation.

These instances form the nodes of a scheduling graph, which is constructed according to the dependencies and ordering constraints specified in the scheduling rules. The graph is rooted in a designated start scenario instance and terminates in one or more end scenario instances, thereby defining a complete execution flow from system initialisation to a target end condition.

Intermediate nodes represent scenario instances whose execution is governed by scheduling rules that define ordering, branching, and dependency relationships. These rules guide the progression of the test execution by determining which scenario instances may follow a given state and under which conditions they are activated. As a result, the scheduling graph encodes not only structural dependencies but also the permissible execution paths through the system under test.

\begin{figure}[H]
	\centering
	\begin{tikzpicture}[
		node distance=1.5cm and 1.5cm,
		every node/.style={draw, circle, minimum size=1cm, font=\small}
		]
		
		\node (start) at (0,4) {$\text{\tt sc}_0$};
		
		\node (i1) at (-3,2) {$\text{\tt sc}_1$};
		\node (i2) at (0,2) {$\text{\tt sc}_2$};
		\node (i3) at (3,2) {$\text{\tt sc}_3$};
		
		\node[draw, circle, double, minimum size=1cm, font=\small] (e1) at (-1.5,0) {$\text{\tt sc}_4$};
		\node[draw, circle, double, minimum size=1cm, font=\small] (e2) at (2,0) {$\text{\tt sc}_5$};
		
		\draw[->] (0,5.1) -- (start);
		\draw[->] (start) -- (i1);
		\draw[->] (start) -- (i2);
		\draw[->] (start) -- (i3);
		
		\draw[->] (i1) -- (e1);
		\draw[->] (i1) -- (i2);
		\draw[->] (i2) -- (e1);
		\draw[->] (i3) -- (e2);
	\end{tikzpicture}
	\caption{Example Scheduling Graph, $\text{\tt sc}_4$ and $\text{\tt sc}_5$ are End-Scenarios.}
	\label{fig:exampleschedgraph}
\end{figure}
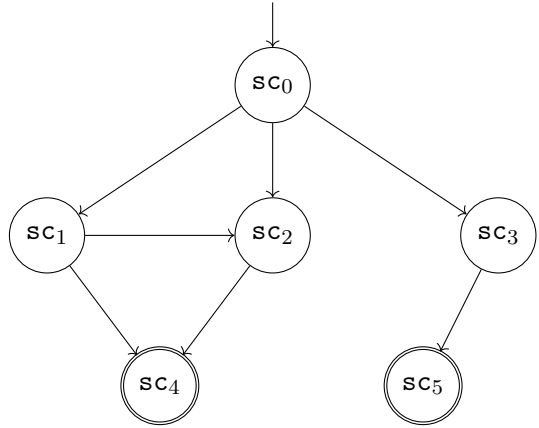
The resulting graph defines how and when individual scenario instances are executed at runtime, thereby enabling the controlled and coordinated execution of multiple interacting scenarios. In this way, the scheduling graph serves as the foundation for end-to-end test generation, ensuring that generated test executions follow well-defined paths from a start scenario to a corresponding end scenario while respecting all specified behavioral and scheduling constraints.

In the third layer, the previously constructed symbolic \buchi automaton and the scheduling graph are processed to derive executable artifacts. This layer comprises two main components: 
\begin{itemize}
	\item test case generation and 
	\item simulation and oracle generation. 
\end{itemize}

This is described in more detail in the next sections.

\subsubsection{Test Generation}
The test generator systematically explores the scheduling graph by traversing all possible paths from the designated start scenario instance to the corresponding end scenario instances. Each path represents a valid end-to-end execution sequence and is therefore considered for test generation. Even paths that differ only in a single transition are explored independently to ensure coverage of all possible execution combinations.

For each path in the scheduling graph, the corresponding sequence of scenario instances is considered. Each scenario instance is represented by a symbolic \buchi automaton. In cases where multiple scenario instances are active concurrently, multiple symbolic \buchi automata are considered in parallel.

Each elementary scenario is additionally associated with its own set of condition/action items, which are preserved during instantiation and therefore also present in the corresponding scenario instances. These condition/action items follow a guarded-command semantics, where execution is defined by the evaluation of a predicate over the current state and scenario variables.

More precisely, each transition is associated with a guard, a condition, and a set of actions that update temporary variables. A transition is enabled if the conjunction of its guard, condition, and the current valuation of temporary variables is satisfied. If this evaluation yields true, the corresponding actions are executed and may update the temporary variable state, thereby influencing subsequent transition evaluations.

This corresponds to a symbolic guarded transition system with state updates, where transitions are defined in the form of guarded commands of the type "if condition then update state". In this work, this semantics is adopted based on the C-Language, where statements are executed sequentially after the condition becomes true, adapted to a symbolic and SMT-solved execution setting. This mechanism allows the system to express control-flow-like behavior within the formal execution model while maintaining compatibility with symbolic evaluation using SMT solving.

Test generation proceeds by exploring the transitions of the involved symbolic \buchi automata starting from their initial states in a depth-first search (DFS) order. Test generation and execution for a given scenario instance only start once its corresponding precondition is satisfied. From that point on, the symbolic Büchi automaton of that instance is traversed in DFS order. If the precondition is not satisfied, the corresponding execution path is not explored and no transitions are traversed for that instance. This mechanism ensures that scenario execution is context-dependent and only activated under valid conditions.

Within active executions, transitions are evaluated based on their guards together with associated condition/action items and the current temporary variable state. The traversal aims to cover relevant behavioral paths, in particular those leading to accepting states. This results in a set of symbolic execution paths characterised by transition guards.

Concrete test cases are obtained by solving the accumulated transition constraints along these paths using Microsoft's Z3\footnote{\url{https://www.microsoft.com/en-us/research/project/z3-3/}} SMT solver. The resulting variable assignments constitute concrete input stimuli for the system under test. Output validation is not embedded in the generated test cases, as correctness is verified separately using runtime oracles.

\begin{figure}[H]
	\centering
	\begin{tikzpicture}[node distance=1.5cm and 1.5cm]
		
		\node[draw, circle, minimum size=1cm, font=\small] (s0) at (2,4) {$\text{\tt s}_0$};
		\node[draw, circle, minimum size=1cm, font=\small] (s1) at (0,2) {$\text{\tt s}_1$};
		\node[draw, circle, double, minimum size=1cm, font=\small] (s3) at (2,0) {$\text{\tt accept}$};
		
		\draw[->] (2,5.1) -- (s0);
		\draw[->] (s0) -- node[midway, left] {$z = 0 \wedge x = p_0 \wedge p_0 = 42$} (s1);
		\draw[->] (s0) -- node[midway, right] {$z = 1 \wedge x = p_0 - 25 \wedge p_0 = 42$} (s3);
		\draw[->] (s1) -- node[midway, left] {$z = p_0 \wedge x = 1 \wedge p_0 = 42$} (s3);
	\end{tikzpicture}
	\caption{Example symbolic \buchi automaton.}
	\label{fig:scsltcgen}
\end{figure}
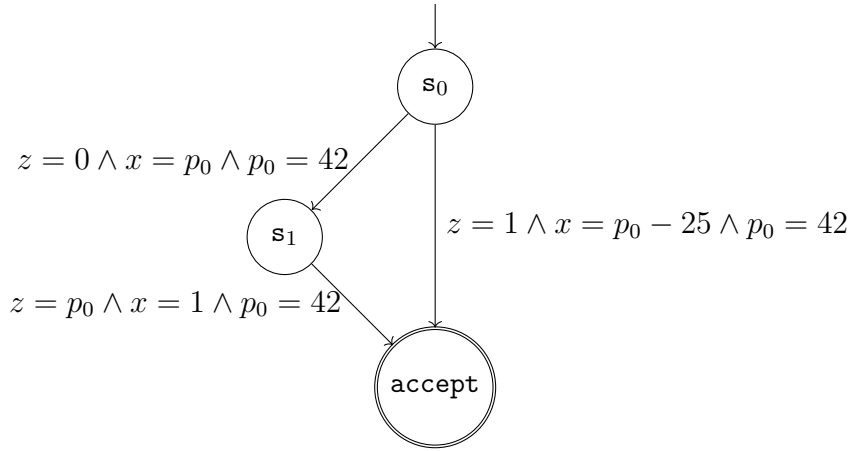

Consider the symbolic \buchi automaton in Figure~\ref{fig:scsltcgen}. Variables $z$ and $x$ are both output variablues while $p_0$ is a parameter of the scenario and set to $p_0 := 42$ in this instance. The test generation algorithm yields a test suite consisting of two test cases, as shown in Table~\ref{tab:tc1} (Test Case 1) and Table~\ref{tab:tc2} (Test Case 2).

\begin{table}[h]
	\centering
	\begin{tabular}{c|c|l|l}
		\textbf{Step} & \textbf{Transition} & \textbf{Transition Expression} & \textbf{Concrete Valuation} \\ \hline
		1 & $s_0 \rightarrow \text{\tt accept}$ & $z = 1 \wedge x = p_0 - 25 \wedge p_0 = 42$ & $\{z:=1, x := 17, p_0 = 42\} $ \\
	\end{tabular}
	\caption{Test Case 1: Direct transition $s_0 \rightarrow \text{\tt accept}$}
	\label{tab:tc1}
\end{table}

\begin{table}[h]
	\centering
	\begin{tabular}{c|c|l|l}
		\textbf{Step} & \textbf{Transition} & \textbf{Transition Expression} & \textbf{Concrete Valuation} \\ \hline
		1 & $s_0 \rightarrow s_1$ & $z = 0 \wedge x = p_0 \wedge p_0 = 42$ & $\{z:=0, x := 42, p_0 = 42\} $ \\
		2 & $s_0 \rightarrow \text{\tt accept}$ & $z = p_0 \wedge x = 1 \wedge p_0 = 42$ & $\{z:=42, x := 1, p_0 = 42\} $ 
	\end{tabular}
	\caption{Test Case 2: Transition $s_0 \rightarrow \text{\tt accept}$}
	\label{tab:tc2}
\end{table}

The resulting test suite is serialised into a structured JSON representation, which enables persistence, interoperability, and execution within the distributed runtime environment. The SUT must pass both test cases of the test suite in order to pass the scenario specification.

\subsubsection{Simulation and Oracle Generation}
Simulation and oracle artifacts are generated by transforming the symbolic \buchi automaton, together with condition/action specifications, into executable \cpp code. Through the Jinja2-based\footnote{\url{https://jinja.palletsprojects.com/en/stable/}} code generation process, each automaton is compiled into a dedicated \cpp class that encapsulates all input variables, constants, and scenario parameters as private member variables. Input variables are updated via external interface calls, while scenario parameters are refreshed at each iteration of the execution loop.

The transitions of the symbolic \buchi automaton are directly mapped to \cpp conditional statements. In particular, each transition guard is translated into an \texttt{if}-condition over the private member variables (e.g., \texttt{x > 0}). If a guard evaluates to true, the corresponding transition is executed, including state updates and optional output actions. Output variables are written via a \texttt{write()} function, which forwards data through a configurable wrapper interface to other agents, enabling flexible deployment across different execution environments. In the cloud-environment, the data is serialised to JSON packages and sent via UDP multicast.

Each automaton state is stored as a private member variable within the generated class and is evaluated inside a cyclic main loop implemented using a switch-case structure.

Importantly, for oracles, the resulting execution model does not require any external constraint solving or runtime SMT reasoning. All transition decisions are resolved through direct evaluation of compiled \cpp boolean expressions, resulting in a lightweight and highly efficient execution mechanism with predictable constant-time guard evaluation.

In contrast, simulation requires the use of an SMT solver to evaluate the outputs of the system under test (SUT) and to determine the subsequent transition to be taken. Furthermore, the SMT solver is used to compute the expected system responses, thereby generating appropriate outputs of the simulation in reaction to the observed SUT behavior. The generated SCSL-Based Simulation utilises the SMT-Solver Microsoft Z3\footnote{\url{https://www.microsoft.com/en-us/research/project/z3-3/}}.

Each oracle or simulation instance follows a uniform execution scheme. First, the corresponding class is instantiated. During runtime, scenario parameters are re-initialised at each iteration of the main loop, ensuring that parameter-dependent behavior remains consistent with the scheduling semantics.

Execution remains idle until the precondition of the specification becomes satisfied. Once activated, the system enters its main evaluation loop. In each iteration, a snapshot of the current input state is taken, and all outgoing transitions of the current active states are evaluated sequentially. If a transition guard evaluates to true, the corresponding action is executed and the automaton transitions to the next state.

Importantly, the nondeterminism inherent to \buchi automata is explicitly preserved within this evaluation scheme. When multiple outgoing transitions are enabled simultaneously, the system does not enforce an arbitrary deterministic choice. Instead, the execution maintains a set of active states, exploring all valid continuation candidates in parallel. If a specific state within this set fails to provide a suitable outgoing transition for the current input, it is pruned. Should the state set become empty, it indicates that the SUT  has deviated from the behavior specified by the test oracle, signaling a violation of the acceptance conditions.

\subsection{Cloud-Based Execution Environment}
The second component of the SCSL Tool Platform is a cloud-based execution environment that orchestrates simulations, oracles, test executions, and the SUT within a unified distributed runtime infrastructure. The environment is designed as a time-synchronised, UDP-based, eventually consistent shared-state communication system for distributed runtime verification.
The environment is designed as a time-synchronised, UDP-based, eventually consistent shared-state communication system for distributed runtime verification, following related approaches in multi-agent and distributed monitoring frameworks~\cite{DBLP:journals/corr/abs-2110-12586,DBLP:journals/corr/abs-2204-09796}.

\subsubsection{Local Agent Execution Model}

Each agent in the execution infrastructure is realised as a self-contained runtime component generated from a symbolic \buchi automaton using a Jinja2-based code generation pipeline. The resulting \cpp{} implementation encapsulates the complete automaton logic, including all input variables, constants, and scenario parameters, as private member variables within a dedicated class. This design ensures strict encapsulation of the execution state and enables deterministic behaviour across distributed deployments.

The generated automaton is executed using a two-threaded runtime model. A dedicated receiver thread is responsible for continuously processing incoming UDP multicast messages. These messages contain JSON-encoded state updates from other agents and are merged into the agent's local input state representation. This thread maintains the most recent view of the distributed system and ensures that the input state is continuously updated asynchronously with respect to the execution logic.

In parallel, an execution thread performs the actual automaton evaluation. At each iteration of a cyclic main loop, the current input state is accessed as a snapshot, ensuring temporal consistency during evaluation. Scenario parameters are re-initialised at the beginning of each loop iteration to reflect the scheduling semantics of the system.

Each agent initially remains in an idle state until its associated precondition becomes satisfied. Once activated, the execution thread continuously evaluates the automaton in a loop, processing transitions based on the latest snapshot of the input state provided by the receiver thread. This separation of communication and execution ensures both responsiveness to distributed state changes and deterministic local behaviour.

\subsubsection{Time-Synchronised Distributed Agent Execution Model}
The execution infrastructure follows a homogeneous agent-based design. Each component, including the SUT, test executor, oracle, and simulation engine, is deployed as an independent agent within a Docker container. All agents adhere to a common execution ontology, meaning that they share a uniform representation of state, time, and message semantics, regardless of their specific role. This enables interchangeable interaction patterns between testing, simulation, and verification components.

Inter-agent communication is implemented using a blackboard-style architecture based on a globally synchronised time tick. At each time step, agents publish their local output state as JSON-encoded messages via UDP multicast over a Docker MACVLAN network. Incoming messages are received asynchronously and merged into each agent's local input state. To ensure deterministic evaluation, all agents take a consistent snapshot of the current input state at each global time tick before performing local computation. 

At each global time tick, the system evolves in a strictly synchronised sequence of operations. First, each agent acquires a snapshot of its current local input state, ensuring that all computations are performed over a consistent view of the distributed system. Based on this snapshot, every agent then evaluates its local transition function, which determines the next control state and the corresponding output actions.

Subsequently, agents emit their output messages via UDP multicast in JSON-encoded form. These messages are disseminated to all other agents participating in the system. Finally, all incoming messages are merged into the respective input states of each agent, forming the basis for the next time step. This cycle defines the discrete-time evolution of the overall distributed system.

At the top level, a dedicated coordinator component is responsible for orchestrating distributed test execution. The coordinator initialises all executor agents using a test execution configuration, which specifies the location of the test suite and binary artifacts in an external object store (e.g., a MinIO-based S3-compatible storage system\footnote{\url{https://www.min.io/}}). Once initialisation is complete, all agents are synchronised and start execution simultaneously at a globally defined start time, using a shared logical clock.

During execution, test cases are transmitted as stimulation inputs by the test executor via UDP multicast in JSON format over the MACVLAN network. The SUT, oracles, and simulations process these inputs uniformly using the same communication and state update mechanisms. This ensures that all runtime components operate on a consistent semantic model of system behaviour.

Throughout the test run, execution status and intermediate verdicts are continuously monitored and displayed in a test management interface. Upon completion, each agent uploads its execution logs and final verdicts to the object store. Finally, all agents are reset to their initial state, enabling reproducible re-execution of test campaigns under identical conditions.

\section{A System Test Example}
The SUT comprises three autonomous rovers tasked with retrieving three items within a bounded environment containing designated exclusion zones. The rovers operate on a two-dimensional Euclidean plane, where the state of each rover $i$ is defined by its coordinates $(x_i, y_i)$. Similarly, each target item $j$ is assigned a static position $(x_j, y_j)$ within the operational area.

\subsection{Test Configuration and Setup}
\begin{figure}[H]
	\centering
	\includegraphics[width=.9\textwidth]{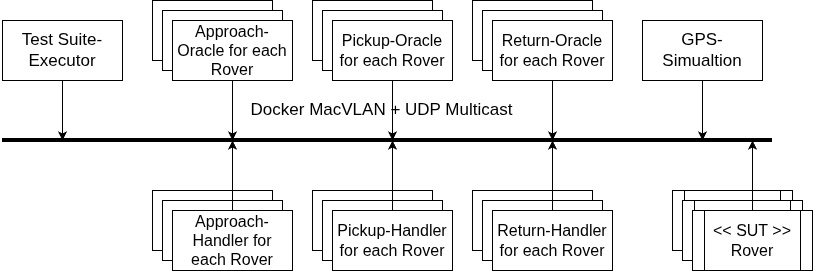}
	\caption{The Architecture of the System Test Execution.}
	\label{fig:schedule}
\end{figure}

The test setup is implemented as a set of communicating Docker containers comprising the simulation components, oracle, test suite execution environment, and the SUT, interconnected via a MACVLAN network and UDP multicast, as illustrated in Figure~\ref{fig:schedule}. The overall test execution is divided into an initialisation phase and an execution phase. During the initialisation phase, items and exclusion zones are distributed across the operational field.

\begin{figure}[H]
	\centering
	\includegraphics[width=.7\textwidth]{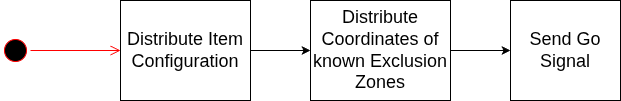}
	\caption{Test Suite Schedule.}
	\label{fig:schedule_ts}
\end{figure}

 An excerpt of the test suite is shown in Listing~\ref{list:tsjson}. These placements are defined as parameterised scenarios within a sequential composition: target items are positioned first, followed by the delineation of exclusion zones, as shown in Figure~\ref{fig:schedule_ts}.

The simulation remains in a paused state until the \texttt{simulation\_start} signal is issued by the \textit{sendSimulationStart} scenario. Upon activation, the \textit{ApproachHandler}, \textit{ReturnHandler}, and \textit{PickupHandler} initiate execution in coordination with all rover-specific scenarios, as governed by the predefined schedule. Throughout the mission, the \textit{Emergent Property Monitor} continuously audits the system state for any property violations.
Listing~\ref{list:tsjson} shows the test configuration in JSON format. This representation is not the original SCSL specification, but an internal representation of the initialisation scenario.

\begin{lstlisting}[basicstyle=\tiny\ttfamily,label=list:tsjson, caption={Excerpt of the Initialization Phase (Scenario 1).}]
	{
		"name": "Scenario-1-Initialization",
		"stimulation": {
			"set_item1": true, "set_item1_x": 5, "set_item1_y": 5,
			"set_item2": true, "set_item2_x": 6, "set_item2_y": 6
		},
		"expected_observations": {
			"condition": ""
		}
	}
\end{lstlisting}

\subsection{Scenario: GPS-Induced Navigation Failure}
To evaluate realistic system behaviour, a GPS signal glitch is simulated. The test is considered passed if at least two of the three rovers survive and all items can be salvaged. This setup allows assessment of the system's response when a rover receives erroneous position data, subsequently enters an exclusion zone, and becomes unavailable. In such cases, the command centre is expected to reassign the task by selecting an alternative rover to collect the respective item.

The scenario is activated under the condition that the rover under consideration is not already located within an exclusion zone and is currently in the Approaching state, i.e. navigating towards a designated item. Once these preconditions are met, the GPS glitch is introduced. Upon reaching the glitch timestamp (i.e., a predefined future time at which a GPS glitch is introduced), the rover's reported position remains static, while the test engine continues to track its true position. As the rover assumes a standstill situation, it continues driving until it enters an exclusion zone. This condition persists until a predefined glitch counter reaches a specified maximum threshold, which is provided as a configurable parameter.

\newpage

\begin{lstlisting}[style=scsl,language=scsl,caption={Scenario for Simulating GPS Glitch.},label=lst:scgpsglitch,mathescape=true]
	elementary scenario GPSGlitch( $r$ : Rover, $\text{\tt glitchTime}$ : $t,\text{\tt glitchCntMax} : \mathbb{R} $ )
	
	spec $\tg \big( (\neg \innga(r.pos,\nga) \wedge r.s = \approaching) {}$
			$\Rightarrow \tx \sactive\big)$;	
	spec  $\tg \big( (\text{\tt glitchCnt} = \text{\tt glitchCntMax}) \Rightarrow  \tx \neg \sactive \big)$
	
	initact $\frm := \{ r.pos \}; \text{\tt glitchCnt} := 0$
	cndact $[t = \text{\tt glitchTime}]/\text{glitchPos} := r.pos$
	cndact $[t > \text{\tt glitchTime}]/r.pos := \text{\tt glitchPos};\text{\tt glitchCnt} = \text{\tt glitchCnt} + 1$
	end scenario      
\end{lstlisting}

This scenario is instantiated with Rover 3. The glitch time is defined to $5$, while the maximum value of the glitch counter is set to $3$. The rover is considered to go to item $(5,7)$ in forward direction, so heading a exclusion zone at $(5,4)$. From the timestamp of 5 seconds after start, its reported position becomes artificially frozen, meaning that further movement is no longer reflected in the received GPS signal, despite the rover potentially continuing its internal navigation process. This discrepancy persists in a time-dependent manner and is regulated by a glitch counter mechanism. The counter increments with continued operation under faulty signal conditions until it reaches the predefined maximum value of $3$. Note that the test engine receives the true position of the rover while the rover receives the faulty position.

Once this limit is reached, the rover continues to assume correct localisation despite the underlying GPS fault. As a result, it believes it remains on its intended trajectory, while its actual position may already be within an exclusion zone that should have been avoided. Due to the faulty trajectory information, this deviation is not detected in time, leading to a violation of the exclusion zone constraint.

\begin{lstlisting}[style=scsl,language=scsl,caption={Scenario Instance for Simulating GPS Glitch 5 seconds after start.},label=lst:scgpsglitchinstance,mathescape=true]
	instance gpsGlitch
	of scenario GPSGlitch( $r$ := $\text{\tt rover3}$, $\text{\tt glitchTime}$ := 5, $\text{\tt glitchCntMax} := 3$ );
\end{lstlisting}

As can be observed in the scenario execution in Listing~\ref{lst:scgpsglitchoutput}, the position of Rover 3 is held constant at $(5,1)$ once the GPS glitch is triggered. Despite this apparent stalling in the reported localisation, the rover remains in the Approaching state throughout this phase, as the high-level behavioural state is not immediately affected by the corrupted position updates.

\newpage 

\begin{lstlisting}[style=bash,language=bash, caption={Output of the System Test with GPS Glitch.}, label=lst:scgpsglitchoutput]
[GPS] Simulate fault for Rover 3
[GPS] Rover 3 True(5,1) GPS(5,1)
[17:28:01.147] [INFO] Rover 3 Pos (5,1) (State: APPROACHING)
[17:28:01.647] [INFO] Rover 3 Pos (5,1) (State: APPROACHING)
[17:28:02.149] [INFO] Rover 3 Pos (5,1) (State: APPROACHING)
[17:28:02.650] [INFO] Rover 3 Pos (5,4) (State: DEAD)
[17:28:07.344] [CommandCentre] Reschedule Item 1 (5, 7) to Rover 2
\end{lstlisting}
The rover continues to operate under the assumption of valid navigation data until the glitch counter reaches its configured maximum value. At this point, normal GPS functionality is restored. However, due to the accumulated discrepancy between the rover's actual trajectory and its previously frozen reported position, the system updates cause an abrupt correction in localisation.

Upon detecting its true position within this exclusion zone, Rover 3 transitions into a failed operational state and actively reports its dead state to the command centre. This notification indicates that the rover is no longer capable of participating in mission execution. Consequently, the system classifies Rover 3 as unavailable for further task allocation, and higher-level coordination logic is triggered to reassign the affected task to an alternative rover (here: Rover 2, as shown in Listing~\ref{lst:scgpsglitchoutput}).

Additionally, it can be observed that, until the end of the system test, the oracle evaluations associated with this rover -- namely the {\it Return} oracle and the {\it Pickup} oracle  -- reach a passed state, as shown in Listing~\ref{lst:gpsglitchoracles}. This can be attributed to the rover's entry into an exclusion zone.

\begin{lstlisting}[style=bash,language=bash, caption={Output of the Oracles during GPS Glitch Test.}, label=lst:gpsglitchoracles]
[Approach3-ORA] PASS.
[Pickup3-ORA] PASS.
[Return3-ORA] PASS.
\end{lstlisting}

The successful oracle resolution can be attributed to the system's correct handling of the rover's disrupted operational lifecycle following the simulated GPS glitch and its transition into the exclusion zone. Despite Rover 3 being unable to complete the original task, the system adapts appropriately, resulting in both the pickup oracle and return oracle reaching a passed state.

As a result, both oracles converge to a successful verification outcome, indicating that the expected behavioural criteria have been satisfied. Furthermore, all items were successfully salvaged and the minimum required number of rovers remained operational. Consequently, the overall system test for Rover 3 is considered passed. The system test took 75 seconds for execution.

\subsection{Parameter Sensitivity}
To evaluate the effectiveness of the \textit{EmergentPropertyChecker} (Listing~\ref{lst:emergentchk}), we conducted a sensitivity analysis by varying the temporal and spatial parameters of the initialization scenarios. 

The property defined in the specification establishes mission-critical deadlines and throughput requirements. By systematically decreasing the parameter $\tdst$ (Time at Destination from start) while increasing the spatial distance of target items, we identified the operational boundaries of the rover swarm.

\begin{table}[h]
	\centering
	\small 
	\caption{Impact of Parameter Variation on Property Verification.}
	\label{tab:parameter_tests}
	\begin{tabular}{@{}llcl@{}}
		\hline
		\textbf{Test ID} & \textbf{Target Dist.} & \textbf{Deadline ($\tdst$)} &  \textbf{Verdict} \\ \hline
		T-1 & 5 steps & 20s  & PASS within assumptions \\
		T-2 & 15 steps & 20s & PASS within assumptions \\
		T-3 & 15 steps & 10s & FAIL: Temporal Violation \\
		T-4 & 25 steps & 20s & FAIL: Temporal Violation \\ \hline
	\end{tabular}
\end{table}

The results demonstrate that the test suite successfully passed under the nominal operational assumptions, where the allotted time $\tdst$ was sufficient for the rovers to traverse the Euclidean distance to the items. Conversely, the framework correctly issued a \textit{FAIL} verdict when the parameters were pushed beyond the physical capabilities of the robots, such as in T-3 and T-4. 

\subsection{Experimental Evaluation}
The experimental evaluation indicates a highly efficient end-to-end workflow for the generation and execution of scenario-based system tests. In particular, the derivation of oracles and simulation artefacts directly from the specification was completed in less than two seconds (Table~\ref{tab:gen_oracle_time}), demonstrating minimal overhead in the transformation step from formal description to executable test components.

\begin{table}[h]
	\centering
	\small 
	\caption{Time to generate executable Oracles and Simulations from SCSL on a AMD Ryzen 7 PRO 5850U CPU.}
	\label{tab:gen_oracle_time}
	\begin{tabular}{ll}
		\hline
		\textbf{Oracle/Simulation} & \textbf{Time} \\ \hline
		Approach Oracle & 0.946s \\
		Pickup Oracle & 0.977s \\
		Return Oracle & 1.622s \\
		Approach Simulation & 0.556s \\
		Pickup Simulation & 0.677s \\
		Return Simulation & 1.142s \\\hline
	\end{tabular}
\end{table}

Furthermore, the generated Oracle and Simulation \cpp code could be compiled without significant delay, and the corresponding runtime environment was set up rapidly. This fast provisioning process enables an iterative testing workflow in which even small modifications to the specification can be translated into new executable tests with negligible turnaround time. As a result, test execution and result retrieval can be performed in a highly responsive manner, supporting rapid experimentation and continuous refinement of scenarios, as shown in Table~\ref{tab:gen_oracle_time}.

In addition, preliminary experiments with simple baseline scenarios revealed a trajectory-following issue in cases where only a single feasible path exists around an exclusion zone. In such situations, the rover fails to correctly navigate along the constrained route, exposing a limitation in the current navigation behaviour. This observation demonstrates that the proposed approach is capable of uncovering non-trivial system faults even in seemingly simple scenarios, thereby highlighting its effectiveness for systematic error detection.

In terms of system communication, the use of Docker-based isolation in combination with MACVLAN networking, and UDP multicast results in a very low communication overhead. This setup allows the distributed components of the test environment to exchange messages efficiently, with minimal latency introduced by the infrastructure layer below 2 ms. Empirical timing observations further suggest that message exchange operates in a near real-time manner, with no noticeable delays or disruption effects under normal test conditions.

Overall, the measurements indicate that the proposed setup is well-suited for time-sensitive system testing, as it achieves both fast test generation and low-latency inter-component communication.

\section{Related Work}\label{sec:related}

\paragraph{Early work in the road vehicle domain}
Gipps~\cite{GIPPS1981105} created a mathematical specification formalism for computer-based car-following simulation.   \emph{Gipps' model} captures the behaviour of a single vehicle in a traffic stream, specifically how it reacts to the vehicle in front. The model specifies each braking and acceleration action of the following vehicle.
Building on Gipps' work, the \emph{Intelligent Driver Model (IDM)}~\cite{Treiber_2000} introduces time-continuous traffic flow scenario specification with environmental aspects. This enables the simulation of braking manoeuvres or minimum spacing between cars. This was further extended by additional physical simulations of longitudinal dynamics, expressed by the \emph{Optimal Velocity Model (OVM)}~\cite{ovm}. OVM enables the dynamical adjustment of the acceleration after the lead vehicle brakes. The model uses mathematical equations to specify acceleration and braking values. 
Ulbrich~et~al. introduced fundamental formal definitions and examples for \emph{scene}, \emph{situation} and \emph{scenario}, where a scene is considered as a snapshot of the environment including static and dynamic elements, a situation is defined as a context-specific scene and a scenario as a description of a temporal evolution of multiple scenes~\cite{DBLP:conf/itsc/UlbrichMRSM15}. Hilscher~et~al. developed an abstract model using spatial interval logic to represent multi-lane roadways with potential lane changes of ego vehicles, called \emph{Multi-Lane Spatial Logic} (MLSL). The formalism enables the verification of lane changes for a single vehicle in an environment with multiple ego vehicles, while a distance controller ensures safe spacing between vehicles~\cite{DBLP:conf/icfem/HilscherLOR11}. Schwammberger~\cite{DBLP:journals/corr/abs-1804-04346} extended MLSL by incorporating property checks into the lane-changing controller for motorway scenarios, ensuring that undesirable events never occur, as initially proposed by Hilscher~et~al.~\cite{DBLP:conf/icfem/HilscherLOR11}. Schwammberger implemented the line-change controller in UPPAAL to verify the implemented timed automaton and the line changing protocol. Damm~et~al. introduced \emph{Traffic Sequence Charts} (TSC) as a formal and visual specification language for designing traffic scenarios in the context of autonomous vehicle testing~\cite{DBLP:conf/birthday/DammMPR18}. Recent work uses TSCs for runtime monitoring of complex scenario-based requirements, thereby supporting scenario-based testing during execution~\cite{Stemmer2025RuntimeMonitoringTSC}. TSCs are designed textually or visually, enabling structure-based validation to improve safety assurance by creating a scenario catalogue.

For scenario specification of distributed event-based systems, \emph{Petri Nets}~\cite{Petri62} are employed to define scenarios in form of subsequent events. Petri Nets   facilitate the formal description of system requirements which are then used as input models for test generation, as outlined in Sarmiento~et~al.~\cite{SARMIENTO2016123}. Today's scenario modelling is primarily graphics-based, like \emph{ASAM OpenSCENARIO}~\cite{asamopenscenario}, which is designed for the automotive domain, focusing on defining real-world traffic scenarios and simulating automotive components. Simulations of traffic behaviours are then utilised to test automotive components.

\paragraph{Recent trends} In the recent literature on scenario-based validation of cyber-physical systems (notably automated driving and, increasingly, automated railway systems), scenario descriptions are used predominantly as artefacts for \emph{testing/validation and runtime monitoring}, and are typically embedded in domain-specific world models and toolchains. This is reflected, for example, by end-to-end scenario-based testing pipelines and scenario generation frameworks for automated driving~\cite{Finkeldei2025ScenarioFactory2,Yan2025OnDemandScenarioGen,Zhao2025LLMSurveyScenarioTesting} as well as closed-loop scenario-based simulation frameworks for highly automated railway systems~\cite{Wild2025RailwayScenarioFramework}.

\paragraph{Wide-spectrum scenario formalisms} In contrast to domain-tailored traffic scenario languages, there also exist wide-spectrum, domain-independent scenario modelling formalisms based on interaction/sequence specifications, for example message sequence charts and their executable variants such as \emph{Live Sequence Charts (LSCs)}. These formalisms are used beyond testing, e.g.\ for requirements specification, consistency checking, and synthesis of reactive behaviour from scenario-based descriptions~\cite{Harel2005StatechartsFromLSC}.

In addition to that, there are several proprietary car traffic simulation platforms, such as \emph{PT-VISSIM}~\cite{PTVISSIM}, which is designed for trajectory modelling.
Beyond automotive applications, the latest \emph{SysML V2} Standard~\cite{sysmlv2} offers syntactic support for domain-independent test scenario definitions, based on occurrences, verification definitions,  constraint definitions, interactions, and state machines. SysML V2 introduces enhancements over its predecessor to better support system-level modelling. The strengths of SysML V2 are in defining system interactions, state-driven behaviour and high-level operational scenarios. 

In the aviation sector, the graphical \emph{Aviation Scenario Definition Language (ASDL)}~\cite{jafer2016formal} supports formal specification of comprehensive aircraft landing scenarios. Scenario specification is also explored through Time Modelling, particularly for system requirements verification. For example, the \emph{Clock Constraint Specification Language (CCSL)}~\cite{DBLP:journals/isse/Mallet08} translates specifications into clock graphs. These graphs facilitate constraint-solving using SMT solvers.

Each existing scenario specification formalism has disadvantages: formalisms like ASAM OpenSCENARIO, IDM, or PT-VISSIM offer strong domain-dependent capabilities for behaviour specification and simulations but lack support for behaviour simulation of arbitrary domains and test generation. On the other hand, SysML V2, Petri-Nets and the Clock Constraint Specification Language (CCSL) are primarily domain-independent, but lack a simple, expressive syntax. Additionally, they are not fully suitable for formal behaviour specification. While test and code generation is possible with these formalisms, it often requires significant resources.

SCSL, however, offers a deliberately simple and expressive, \emph{scenario-centric} syntax for formal system-test scenario specification. By leveraging existing formalisms from UML/SysML and VDM, the LTL syntax is extended by using condition-actions on state transitions and frames to restrict variable modifications. SCSL is intentionally designed to be domain-independent, supporting specification of re-usable, parametrised scenario models that can be   assembled into domain-specific libraries. The state-based behaviour of LTL is a notable feature that enables straightforward test and code generation from specified scenarios.

\section{Discussion}\label{sec:conc}
 
\subsection{Conclusion}

We have introduced the domain-independent scenario specification language SCSL and illustrated its use for system test specifications verifying collaborating robots. Compared to other existing scenario modelling languages, in particular, those based on variants of UML and SysML, SCSL has a fairly simple syntax and a formal behavioural semantics that can be comprehensively specified with acceptable complexity in a compact way. 
  
Domain-specific support can be easily realised by introducing libraries of predefined types, objects, and elementary scenarios, the latter specifying typical behaviours of the application domain, such as physical laws applicable under specific environmental conditions or typical manoeuvres of the domain-specific objects. The usability of SCSL has been demonstrated by means of a system test example describing a salvage mission of collaborating robots.

A major advantage of the formal SCSL semantics is that SCSL specifications can be automatically transformed into test data generators, executable simulations or into test oracles to be embedded into original equipment harnesses stimulating a system component under test and checking its reactions against SCSL specifications.
 
\subsection{Future Work} 

While this paper focuses on introducing the syntax and semantics of SCSL, 
the suitability and effectiveness of SCSL will be demonstrated in a case study using
specifications from the EULYNX standard\footnote{\url{https://eulynx.eu/resource-hub-deliverables/}} that has been designed for harmonising the communication between components of modern railway signalling systems in Europe. We will compare the effectiveness of SCSL to that of well-known model-based testing approaches that rely on comprehensive monolithic system models. 
Relevant evaluation criteria include modelling effort, reusability, scalability, requirements coverage, and fault detection capability. For example, mutation testing could be applied to assess the relative fault detection power of both approaches. In domains such as automotive or avionics, where MBT is well established, comparative studies will help determine whether scenario-based specifications offer measurable advantages in terms of productivity, maintainability, or test completeness. To the best of our knowledge, there is currently no comprehensive and domain-independent empirical study systematically comparing the effectiveness of scenario-based testing approaches—such as those enabled by SCSL—with classical model-based testing approaches relying on monolithic behavioural models. As such, we consider this a novel and promising direction for future work.

To facilitate the creation of LTL specifications that are essential for describing the behaviour of a scenario, we plan to use LTL-specific generalised pre-trained transformers (GPTs) that are capable of creating LTL formulae from natural language specifications~\cite{DBLP:conf/emnlp/ChenGZF23}. A recent publication by Peled et al.~\cite{DBLP:conf/vecos/CohenP23a} indicates that this can be achieved with a very low failure rate. The automated generation can be complemented by   mechanisms to detect erroneous LTL encodings with high probability.\footnote{Suitable validation mechanisms range from automated procedures to others involving manual interaction. An example for automated validation is to let a GPT create witness traces for the created formula and its negation. Then it is checked by a verified algorithm that these witnesses indeed fulfil the formulae. An error in witness generation indicates that the GPT does not have the ``correct understanding'' of the formula semantics, so the text-to-formula transformation cannot be trusted. An example for validation involving manual interaction is to let engineers decide whether correct formula witnesses conform to their expectations expressed in the natural language specification. This manual process can be supported by letting a GPT transform witness traces of even the whole generated formula back to natural language.}

As readers may have noticed, the essential feature of behavioural SCSL scenario specifications is that they can be interpreted on sequences of valuation functions whose values have been obtained from components operating with cyclic execution semantics. Consequently, the combined LTL and condition-action syntax is not the only way to write behavioural SCSL specifications. As  an alternative, for example, RoboSim state machines\footnote{\url{https://www.cs.york.ac.uk/circus/RoboCalc/robosim/robosim-reference.pdf}} \cite{CAVALCANTI20191}  and similar discrete time formalisms could be used.




\newpage
\bibliographystyle{elsarticle-num}
\bibliography{jp, hidyve, fuzzing_refs, references}

\begin{thebibliography}{10}
\expandafter\ifx\csname url\endcsname\relax
  \def\url#1{\texttt{#1}}\fi
\expandafter\ifx\csname urlprefix\endcsname\relax\def\urlprefix{URL }\fi
\expandafter\ifx\csname href\endcsname\relax
  \def\href#1#2{#2} \def\path#1{#1}\fi

\bibitem{DBLP:conf/isola/0001BH18}
J.~Peleska, J.~Brauer, W.~Huang,
  \href{https://doi.org/10.1007/978-3-030-03427-6\_11}{Model-based testing for
  avionic systems -- proven benefits and further challenges}, in: T.~Margaria,
  B.~Steffen (Eds.), Leveraging Applications of Formal Methods, Verification
  and Validation. Industrial Practice - 8th International Symposium, ISoLA
  2018, Limassol, Cyprus, November 5-9, 2018, Proceedings, Part {IV}, Vol.
  11247 of Lecture Notes in Computer Science, Springer, 2018, pp. 82--103.
\newblock \href {https://doi.org/10.1007/978-3-030-03427-6\_11}
  {\path{doi:10.1007/978-3-030-03427-6\_11}}.
\newline\urlprefix\url{https://doi.org/10.1007/978-3-030-03427-6\_11}

\bibitem{DBLP:conf/itsc/HauerSHP19}
F.~Hauer, T.~Schmidt, B.~Holzm{\"{u}}ller, A.~Pretschner,
  \href{https://doi.org/10.1109/ITSC.2019.8917326}{Did we test all scenarios
  for automated and autonomous driving systems?}, in: 2019 {IEEE} Intelligent
  Transportation Systems Conference, {ITSC} 2019, Auckland, New Zealand,
  October 27-30, 2019, {IEEE}, 2019, pp. 2950--2955.
\newblock \href {https://doi.org/10.1109/ITSC.2019.8917326}
  {\path{doi:10.1109/ITSC.2019.8917326}}.
\newline\urlprefix\url{https://doi.org/10.1109/ITSC.2019.8917326}

\bibitem{10.1007/978-3-540-31848-4_6}
K.~G. Larsen, M.~Mikucionis, B.~Nielsen, Online testing of real-time systems
  using uppaal, in: J.~Grabowski, B.~Nielsen (Eds.), Formal Approaches to
  Software Testing, Springer Berlin Heidelberg, Berlin, Heidelberg, 2005, pp.
  79--94.

\bibitem{DBLP:journals/dc/BrookesR91}
S.~D. Brookes, A.~W. Roscoe, \href{https://doi.org/10.1007/BF01784721}{Deadlock
  analysis in networks of communicating processes}, Distributed Comput. 4
  (1991) 209--230.
\newblock \href {https://doi.org/10.1007/BF01784721}
  {\path{doi:10.1007/BF01784721}}.
\newline\urlprefix\url{https://doi.org/10.1007/BF01784721}

\bibitem{DBLP:journals/csur/NielsenLFWP15}
C.~B. Nielsen, P.~G. Larsen, J.~S. Fitzgerald, J.~Woodcock, J.~Peleska,
  \href{https://doi.org/10.1145/2794381}{Systems of systems engineering: Basic
  concepts, model-based techniques, and research directions}, {ACM} Comput.
  Surv. 48~(2) (2015) 18:1--18:41.
\newblock \href {https://doi.org/10.1145/2794381} {\path{doi:10.1145/2794381}}.
\newline\urlprefix\url{https://doi.org/10.1145/2794381}

\bibitem{DBLP:conf/taros/PeleskaBHH25}
J.~Peleska, F.~Br{\"{u}}ning, A.~E. Haxthausen, W.~Huang,
  \href{https://doi.org/10.1007/978-3-032-01486-3\_25}{Scenario-based system
  testing for distributed robotics applications - (invited paper)}, in:
  A.~Cavalcanti, S.~Foster, R.~Richardson (Eds.), Towards Autonomous Robotic
  Systems - 26th Annual Conference, {TAROS} 2025, York, UK, August 20-22, 2025,
  Proceedings, Vol. 16045 of Lecture Notes in Computer Science, Springer, 2025,
  pp. 310--337.
\newblock \href {https://doi.org/10.1007/978-3-032-01486-3\_25}
  {\path{doi:10.1007/978-3-032-01486-3\_25}}.
\newline\urlprefix\url{https://doi.org/10.1007/978-3-032-01486-3\_25}

\bibitem{DBLP:conf/iceccs/CavalcantiH23}
A.~Cavalcanti, R.~M. Hierons,
  \href{https://doi.org/10.1109/ICECCS59891.2023.00010}{Challenges in testing
  of cyclic systems}, in: Y.~A{\"{\i}}t{-}Ameur, F.~Khendek, D.~M{\'{e}}ry
  (Eds.), 27th International Conference on Engineering of Complex Computer
  Systems, {ICECCS} 2023, Toulouse, France, June 14-16, 2023, {IEEE}, 2023, pp.
  1--6.
\newblock \href {https://doi.org/10.1109/ICECCS59891.2023.00010}
  {\path{doi:10.1109/ICECCS59891.2023.00010}}.
\newline\urlprefix\url{https://doi.org/10.1109/ICECCS59891.2023.00010}

\bibitem{DBLP:conf/icfem/HilscherLOR11}
M.~Hilscher, S.~Linker, E.~Olderog, A.~P. Ravn,
  \href{https://doi.org/10.1007/978-3-642-24559-6\_28}{An abstract model for
  proving safety of multi-lane traffic manoeuvres}, in: S.~Qin, Z.~Qiu (Eds.),
  Formal Methods and Software Engineering - 13th International Conference on
  Formal Engineering Methods, {ICFEM} 2011, Durham, UK, October 26-28, 2011.
  Proceedings, Vol. 6991 of Lecture Notes in Computer Science, Springer, 2011,
  pp. 404--419.
\newblock \href {https://doi.org/10.1007/978-3-642-24559-6\_28}
  {\path{doi:10.1007/978-3-642-24559-6\_28}}.
\newline\urlprefix\url{https://doi.org/10.1007/978-3-642-24559-6\_28}

\bibitem{spivey1992znotation}
J.~M. Spivey, The {Z} Notation: A Reference Manual, 2nd Edition, Prentice Hall,
  New York, NY, USA, 1992.

\bibitem{jonesVDM}
C.~B. Jones, Systematic Software Development Using VDM, Prentice-Hall, 1986.

\bibitem{uml_2_5}
{Object Management Group}, {OMG Unified Modeling Language (OMG UML)},
  superstructure, version 2.5.1, Tech. rep., {OMG} (2017).

\bibitem{10.1145/356625.356626}
I.~E. Sutherland, R.~F. Sproull, R.~A. Schumacker,
  \href{https://doi.org/10.1145/356625.356626}{A characterization of ten
  hidden-surface algorithms}, ACM Comput. Surv. 6~(1) (1974) 1–55.
\newblock \href {https://doi.org/10.1145/356625.356626}
  {\path{doi:10.1145/356625.356626}}.
\newline\urlprefix\url{https://doi.org/10.1145/356625.356626}

\bibitem{10.1007/978-3-319-11737-9_24}
S.~Mochizuki, M.~Shimakawa, S.~Hagihara, N.~Yonezaki, Fast translation from ltl
  to b{\"u}chi automata via non-transition-based automata, in: S.~Merz, J.~Pang
  (Eds.), Formal Methods and Software Engineering, Springer International
  Publishing, Cham, 2014, pp. 364--379.

\bibitem{DBLP:journals/corr/abs-2110-12586}
K.~I. Eder, W.~Huang, J.~Peleska,
  \href{https://doi.org/10.4204/EPTCS.348.4}{Complete agent-driven model-based
  system testing for autonomous systems}, in: M.~Farrell, M.~Luckcuck (Eds.),
  Proceedings Third Workshop on Formal Methods for Autonomous Systems, {FMAS}
  2021, Virtual, October 21-22, 2021, {EPTCS}, 2021, pp. 54--72.
\newblock \href {https://doi.org/10.4204/EPTCS.348.4}
  {\path{doi:10.4204/EPTCS.348.4}}.
\newline\urlprefix\url{https://doi.org/10.4204/EPTCS.348.4}

\bibitem{DBLP:journals/corr/abs-2204-09796}
R.~Ganguly, Y.~Xue, A.~Jonckheere, P.~Ljung, B.~Schornstein, B.~Bonakdarpour,
  M.~Herlihy, \href{https://doi.org/10.48550/arXiv.2204.09796}{Distributed
  runtime verification of metric temporal properties for cross-chain
  protocols}, CoRR abs/2204.09796 (2022).
\newblock \href {http://arxiv.org/abs/2204.09796} {\path{arXiv:2204.09796}},
  \href {https://doi.org/10.48550/ARXIV.2204.09796}
  {\path{doi:10.48550/ARXIV.2204.09796}}.
\newline\urlprefix\url{https://doi.org/10.48550/arXiv.2204.09796}

\bibitem{GIPPS1981105}
P.~Gipps,
  \href{https://www.sciencedirect.com/science/article/pii/0191261581900370}{A
  behavioural car-following model for computer simulation}, Transportation
  Research Part B: Methodological 15~(2) (1981) 105--111.
\newblock \href {https://doi.org/https://doi.org/10.1016/0191-2615(81)90037-0}
  {\path{doi:https://doi.org/10.1016/0191-2615(81)90037-0}}.
\newline\urlprefix\url{https://www.sciencedirect.com/science/article/pii/0191261581900370}

\bibitem{Treiber_2000}
M.~Treiber, A.~Hennecke, D.~Helbing,
  \href{http://dx.doi.org/10.1103/PhysRevE.62.1805}{Congested traffic states in
  empirical observations and microscopic simulations}, Physical Review E 62~(2)
  (2000) 1805–1824.
\newblock \href {https://doi.org/10.1103/physreve.62.1805}
  {\path{doi:10.1103/physreve.62.1805}}.
\newline\urlprefix\url{http://dx.doi.org/10.1103/PhysRevE.62.1805}

\bibitem{ovm}
M.~Bando, K.~Hasebe, K.~Nakanishi, A.~Nakayama,
  \href{https://link.aps.org/doi/10.1103/PhysRevE.58.5429}{Analysis of optimal
  velocity model with explicit delay}, Phys. Rev. E 58 (1998) 5429--5435.
\newblock \href {https://doi.org/10.1103/PhysRevE.58.5429}
  {\path{doi:10.1103/PhysRevE.58.5429}}.
\newline\urlprefix\url{https://link.aps.org/doi/10.1103/PhysRevE.58.5429}

\bibitem{DBLP:conf/itsc/UlbrichMRSM15}
S.~Ulbrich, T.~Menzel, A.~Reschka, F.~Schuldt, M.~Maurer,
  \href{https://doi.org/10.1109/ITSC.2015.164}{Defining and substantiating the
  terms scene, situation, and scenario for automated driving}, in: {IEEE} 18th
  International Conference on Intelligent Transportation Systems, {ITSC} 2015,
  Gran Canaria, Spain, September 15-18, 2015, {IEEE}, 2015, pp. 982--988.
\newblock \href {https://doi.org/10.1109/ITSC.2015.164}
  {\path{doi:10.1109/ITSC.2015.164}}.
\newline\urlprefix\url{https://doi.org/10.1109/ITSC.2015.164}

\bibitem{DBLP:journals/corr/abs-1804-04346}
M.~Schwammberger, \href{https://doi.org/10.4204/EPTCS.269.3}{Introducing
  liveness into multi-lane spatial logic lane change controllers using
  {UPPAAL}}, in: M.~Gleirscher, S.~Kugele, S.~Linker (Eds.), Proceedings 2nd
  International Workshop on Safe Control of Autonomous Vehicles, SCAV@CPSWeek
  2018, Porto, Portugal, 10th April 2018, Vol. 269 of {EPTCS}, 2018, pp.
  17--31.
\newblock \href {https://doi.org/10.4204/EPTCS.269.3}
  {\path{doi:10.4204/EPTCS.269.3}}.
\newline\urlprefix\url{https://doi.org/10.4204/EPTCS.269.3}

\bibitem{DBLP:conf/birthday/DammMPR18}
W.~Damm, E.~M{\"{o}}hlmann, T.~Peikenkamp, A.~Rakow,
  \href{https://doi.org/10.1007/978-3-319-95246-8\_11}{A formal semantics for
  traffic sequence charts}, in: M.~Lohstroh, P.~Derler, M.~Sirjani (Eds.),
  Principles of Modeling - Essays Dedicated to Edward A. Lee on the Occasion of
  His 60th Birthday, Vol. 10760 of Lecture Notes in Computer Science, Springer,
  2018, pp. 182--205.
\newblock \href {https://doi.org/10.1007/978-3-319-95246-8\_11}
  {\path{doi:10.1007/978-3-319-95246-8\_11}}.
\newline\urlprefix\url{https://doi.org/10.1007/978-3-319-95246-8\_11}

\bibitem{Stemmer2025RuntimeMonitoringTSC}
R.~Stemmer, I.~Saxena, L.~Panneke, D.~Grundt, A.~Austel, E.~M{\"o}hlmann,
  B.~Westphal, Runtime monitoring of complex scenario-based requirements for
  autonomous driving functions, Science of Computer Programming 244 (2025)
  103301.
\newblock \href {https://doi.org/10.1016/j.scico.2025.103301}
  {\path{doi:10.1016/j.scico.2025.103301}}.

\bibitem{Petri62}
C.~A. Petri, {Kommunikation mit Automaten}, Dissertation, Schriften des IIM~2,
  Rheinisch-Westf{\"a}lisches Institut f{\"u}r Instrumentelle Mathematik an der
  Universit{\"a}t Bonn, Bonn (1962).

\bibitem{SARMIENTO2016123}
E.~Sarmiento, J.~C. Leite, E.~Almentero, G.~{Sotomayor Alzamora},
  \href{https://www.sciencedirect.com/science/article/pii/S1571066116301153}{Test
  scenario generation from natural language requirements descriptions based on
  petri-nets}, Electronic Notes in Theoretical Computer Science 329 (2016)
  123--148, cLEI 2016 - The Latin American Computing Conference.
\newblock \href {https://doi.org/https://doi.org/10.1016/j.entcs.2016.12.008}
  {\path{doi:https://doi.org/10.1016/j.entcs.2016.12.008}}.
\newline\urlprefix\url{https://www.sciencedirect.com/science/article/pii/S1571066116301153}

\bibitem{asamopenscenario}
{ASAM}, {ASAM OpenSCENARIO},
  \url{https://publications.pages.asam.net/standards/ASAM_OpenSCENARIO/ASAM_OpenSCENARIO_XML/latest/index.html}
  (2024).

\bibitem{Finkeldei2025ScenarioFactory2}
F.~Finkeldei, C.~Thees, J.-N. Weghorn, M.~Althoff, Scenario factory 2.0:
  Scenario-based testing of automated vehicles with commonroad, Automotive
  Innovation 8 (2025) 207--220.
\newblock \href {https://doi.org/10.1007/s42154-025-00360-0}
  {\path{doi:10.1007/s42154-025-00360-0}}.

\bibitem{Yan2025OnDemandScenarioGen}
S.~Yan, X.~Zhang, K.~Hao, H.~Xin, Y.~Luo, J.~Yang, M.~Fan, C.~Yang, J.~Sun,
  Z.~Yang, On-demand scenario generation for testing automated driving systems,
  2025, accepted by FSE 2025.
\newblock \href {https://doi.org/10.1145/3715722} {\path{doi:10.1145/3715722}}.

\bibitem{Zhao2025LLMSurveyScenarioTesting}
Y.~Zhao, J.~Zhou, D.~Bi, T.~Mihalj, J.~Hu, A.~Eichberger, A survey on the
  application of large language models in scenario-based testing of automated
  driving systems, IEEE Transactions on Intelligent Transportation Systems
  (2025).
\newblock \href {https://doi.org/10.48550/arXiv.2505.16587}
  {\path{doi:10.48550/arXiv.2505.16587}}.

\bibitem{Wild2025RailwayScenarioFramework}
M.~Wild, J.~S. Becker, C.~Schneiders, E.~M{\"o}hlmann, A scenario-based
  simulation framework for testing of highly automated railway systems, in:
  Proceedings of the 11th International Conference on Vehicle Technology and
  Intelligent Transport Systems (VEHITS 2025), 2025, pp. 88--99.
\newblock \href {https://doi.org/10.5220/0013286600003941}
  {\path{doi:10.5220/0013286600003941}}.

\bibitem{Harel2005StatechartsFromLSC}
D.~Harel, H.~Kugler, A.~Pnueli,
  \href{https://www.microsoft.com/en-us/research/wp-content/uploads/2005/01/he05.pdf}{Synthesis
  revisited: Generating statechart models from scenario-based requirements},
  2005, microsoft Research Technical Report / preprint.
\newline\urlprefix\url{https://www.microsoft.com/en-us/research/wp-content/uploads/2005/01/he05.pdf}

\bibitem{PTVISSIM}
{PTV Group}, Multimodal traffic simulation software,
  \url{https://www.ptvgroup.com/en/products/ptv-vissim} (2024).

\bibitem{sysmlv2}
{OMG}, {OMG} systems modeling language v2,
  \url{https://www.omg.org/spec/SysML/2.0/Beta2/About-SysML} (2024).

\bibitem{jafer2016formal}
S.~Jafer, B.~Chhaya, U.~Durak, T.~Gerlach, Formal scenario definition language
  for aviation: aircraft landing case study, in: AIAA modeling and simulation
  technologies conference, 2016, p. 3521.

\bibitem{DBLP:journals/isse/Mallet08}
F.~Mallet, \href{https://doi.org/10.1007/s11334-008-0055-2}{Clock constraint
  specification language: specifying clock constraints with {UML/MARTE}},
  Innov. Syst. Softw. Eng. 4~(3) (2008) 309--314.
\newblock \href {https://doi.org/10.1007/S11334-008-0055-2}
  {\path{doi:10.1007/S11334-008-0055-2}}.
\newline\urlprefix\url{https://doi.org/10.1007/s11334-008-0055-2}

\bibitem{DBLP:conf/emnlp/ChenGZF23}
Y.~Chen, R.~Gandhi, Y.~Zhang, C.~Fan,
  \href{https://doi.org/10.18653/v1/2023.emnlp-main.985}{{NL2TL:} transforming
  natural languages to temporal logics using large language models}, in:
  H.~Bouamor, J.~Pino, K.~Bali (Eds.), Proceedings of the 2023 Conference on
  Empirical Methods in Natural Language Processing, {EMNLP} 2023, Singapore,
  December 6-10, 2023, Association for Computational Linguistics, 2023, pp.
  15880--15903.
\newblock \href {https://doi.org/10.18653/V1/2023.EMNLP-MAIN.985}
  {\path{doi:10.18653/V1/2023.EMNLP-MAIN.985}}.
\newline\urlprefix\url{https://doi.org/10.18653/v1/2023.emnlp-main.985}

\bibitem{DBLP:conf/vecos/CohenP23a}
I.~Cohen, D.~Peled,
  \href{https://doi.org/10.1007/978-3-031-73741-1\_23}{End-to-end {AI}
  generated runtime verification from natural language specification}, in:
  B.~Steffen (Ed.), Bridging the Gap Between {AI} and Reality - First
  International Conference, AISoLA 2023, Crete, Greece, October 23-28, 2023,
  Selected Papers, Vol. 14129 of Lecture Notes in Computer Science, Springer,
  2023, pp. 362--384.
\newblock \href {https://doi.org/10.1007/978-3-031-73741-1\_23}
  {\path{doi:10.1007/978-3-031-73741-1\_23}}.
\newline\urlprefix\url{https://doi.org/10.1007/978-3-031-73741-1\_23}

\bibitem{CAVALCANTI20191}
A.~Cavalcanti, A.~Sampaio, A.~Miyazawa, P.~Ribeiro, M.~{Conserva Filho},
  A.~Didier, W.~Li, J.~Timmis,
  \href{https://www.sciencedirect.com/science/article/pii/S0167642318301655}{Verified
  simulation for robotics}, Science of Computer Programming 174 (2019) 1--37.
\newblock \href {https://doi.org/https://doi.org/10.1016/j.scico.2019.01.004}
  {\path{doi:https://doi.org/10.1016/j.scico.2019.01.004}}.
\newline\urlprefix\url{https://www.sciencedirect.com/science/article/pii/S0167642318301655}

\bibitem{DBLP:conf/ijcai/GiacomoV13}
G.~D. Giacomo, M.~Y. Vardi,
  \href{http://www.aaai.org/ocs/index.php/IJCAI/IJCAI13/paper/view/6997}{Linear
  temporal logic and linear dynamic logic on finite traces}, in: F.~Rossi
  (Ed.), {IJCAI} 2013, Proceedings of the 23rd International Joint Conference
  on Artificial Intelligence, Beijing, China, August 3-9, 2013, {IJCAI/AAAI},
  2013, pp. 854--860.
\newline\urlprefix\url{http://www.aaai.org/ocs/index.php/IJCAI/IJCAI13/paper/view/6997}

\bibitem{apt2010}
K.~R. Apt, F.~S. de~Boer, E.-R. Olderog, Verification of Sequential and
  Concurrent Programs, Springer, Berlin Heidelberg New York, 2010.

\end{thebibliography}
\newpage
\appendix
\section{Behavioural Semantics of SCSL\\ System Test Configurations}\label{sec:scslsem}

\subsection{Valuation functions} The behaviour of system test configurations is   formalised by finite sequences $\pi = \sigma_0.\sigma_1\dots\sigma_q$ of valuation functions $\sigma_i : V_i \fun D$. 
The index $i$ of valuation $\sigma_i$ is the number of the observation and stimulation step performed by the test equipment, and $\sigma_i(v)$ is the value of some symbol $v\in V_i$ that has been set or observed by the test equipment in this step.
The symbol sets $V_i$ contain the following variable names.
\begin{enumerate}
\item All parameter symbols of objects that are active in step $i$. If, for example, rover number $j$ is part of the collaboration and active in step $i$, then $V_i$ contains symbols $coll.r[i].cmd$, $coll.r[i].dst$, $coll.r[i].id$, $coll.r[i].pos$, $coll.r[i].id$.

\item All auxiliary variable symbols of active instances of elementary scenarios, including the implicitly defined auxiliary variables like $\sactive$. If scenario instance \textsl{Return}($coll.r[j]$) is active in step $i$, for example, $V_i$ contains symbols $\textsl{Return}($coll.r[j]$).\sactive$ and $\textsl{Return}($coll.r[j]$).aux_\text{isLoaded}$.

\item The global auxiliary variable symbol $\tend$ indicating the end of the system test.
\end{enumerate}
The set $D$ is the union of all variable types.  The symbol sets $V_i$ change between steps under the following conditions.
\begin{enumerate}
\item If an object is deleted from the collaboration in step $i$, its parameter symbols are no longer present in $V_{i+1}$.
\item If an object is added to the collaboration in step $i$, its parameter symbols   become elements of $V_{i+1}, V_{i+2},\dots$ until the object is removed again from the collaboration.
\item If an elementary scenario instance is runnable and its precondition is fulfilled in step $i$, its auxiliary variables are visible in $V_{i+1}$.
\item If the execution state of an elementary scenario instance changes from $\sactive$ to  $\neg\sactive$ in step $i$, then its auxiliary variable symbols are no longer contained in $V_{i+1}$.

\item A schedule is illegal if an active scenario tries to evaluate attributes of a non-existing object (``runtime error''). Active scenarios can always check $\text{\textsl{object reference}} \neq \scnull$ to ensure that these runtime errors do not occur.

\end{enumerate}

Symbol $\tend$ is contained in all symbol sets $V_0, V_1, \dots, V_q$. For  valuations $\sigma_i, \ i = 0,\dots,(q-1)$, its value is $\sigma_i(\tend) = \fff$. Only the last valuation fulfils $\sigma_q(\tend) = \ttt$.

\subsection{Execution cycles.} 
Depending on the object type and its \textsf{cycletime}, objects    need $k \ge 1$ observation steps of the test equipment to complete one execution cycle.  During this processing time, outputs remain unchanged, and only the last write to an input is considered in the next processing cycle. For example, if   a processing cycle of the object starts in observation step $i$, the input values specified by $\sigma_i$ are used in this cycle. Any output parameter $y$ remains unchanged until the end of the cycle, that is, $\sigma_i(y) = \sigma_{i+1}(y) = \dots = \sigma_{i+k-1}(y)$, and   $\sigma_{i+k}(y)$ returns the new value of $y$.  If input parameter $z$ is an end point of an interface connecting output $y$ with $z$, then
$\sigma_{i+1}(z) = \sigma_{i+2}(z) = \dots = \sigma_{i+k}(z) = \sigma_i(y)$ and $\sigma_{i+k+1}(z) = \sigma_{i+k}(y)$.
For the next processing cycle starting in step $(i+k)$ and any input parameter $x$, the valuation $\sigma_{i+k}(x)$ is considered. Previous valuations $\sigma_{i+k-1}(x),\dots,\sigma_{i+1}(x)$ are disregarded.

\subsection{Scheduling.} The following scheduling rules determine when an elementary scenario instance becomes active.
\begin{enumerate}
\item Every scenario instance transits through execution states 
$$
\textsl{passive} \trans \textsl{runnable} \trans \textsl{active} \trans \textsl{passive}.
$$
\item State active is characterised for a scenario instance $S$ by $S.\sactive = \ttt$.
\item The auxiliary symbols of a scenario are visible in $V_i$ if and only if it is active in observation step $i$. 
\item A runnable scenario instance becomes active in step $i+1$, if its precondition  $\varphi$ evaluates to $\ttt$ in step $i$. Since   $\varphi$  only refers to object parameters and never to auxiliary variables, it can be evaluated in any step $i$ where all object parameter symbols occurring in $\varphi$ are contained in $V_i$. If this is not the case because parameters of objects are referenced that are no longer part of the collaboration, the precondition is considered to be $\fff$.
\item In a parallel composition of scenario instances $S_1\parallel S_2$, both $S_1$ and $S_2$ become immediately runnable.
\item In a sequential   composition of scenario instances $S_1;S_2$,   $S_2$ remains passive until the $S_1$ has traversed execution states $\textsl{passive} \trans \textsl{runnable} \trans \textsl{active} \trans \textsl{passive}$, whereafter $S_2$ becomes runnable.
\end{enumerate}

\subsection{The effect of interface definitions}
Suppose that
   a collaboration has interface $$\textsf{interface}\ \mathit{I}\ \textsf{from}\ O_1.a\ \textsf{to}\  O_2.b$$
   with objects $O_1, O_2$.
An interface transfers the data ``as fast as possible'', that is, in one observation cycle, from source to target. Therefore, in any     
test execution $\pi = \sigma_0.\sigma_1\dots\sigma_q$, the valuations involved satisfy
$$
\begin{array}{l}
    \forall i:0..(q-1) \centerdot O_1.\sactive \in \dom\sigma_i \wedge   O_2.\sactive \in \dom\sigma_{i+1} \wedge  {} 
    \\     \hspace*{20mm} 
    \sigma_i(O_1.\sactive) \wedge \sigma_{i+1}(O_2.\sactive) \Rightarrow 
    \sigma_{i+1}(O_2.b) =  \sigma_i(O_1.a).
    \end{array}
$$

\subsection{The effect of collaboration changes}

As described in Section~\ref{sec:systestconfig} dynamic collaboration changes are triggered by operations $\mathsf{delete}$ and $\mathsf{create}$ to be called in an action of an elementary scenario instance. This changes the collaboration variable in a atomic way: concurrent calls to change the collaboration are processed in an interleaving manner with nondeterministic sequencing. The collaboration acts like a \emph{monitor} as created, for example, in Java using the \textsl{synchronized} statement. As with writes to other auxiliary variables written to in an action at cycle~$i$, the effect of the collaboration change becomes visible at the beginning of the next cycle $i+1$, so that  the data distribution performed in cycle $i+1$ is already based on the new collaboration status with its new interface wiring.

\subsection{Execution semantics of elementary scenario instances}

Given an elementary scenario instance $S$ and a system test execution $\pi = \sigma_0.\sigma_1\dots\sigma_q$, the execution \emph{$\pi$ conforms to $S$} if  certain   $S$-specific rules hold. To explain these rules, let $\pi'$ be the trace segment of $\pi$ where $S$ is active. Let $\Phi$ be the conjunction of LTL formulae that have been specified in the elementary scenario  type of $S$, but with all parameter symbols exchanged by the corresponding concrete object parameters. 

For example, scenario type \textsl{Pickup} refers to symbols $r, cc.cmd[i], cc.id[i]$ and others. These symbols are exchanged in $\Phi$ by the concrete object references and parameter names that are specified in the system test collaboration, that is, 
$coll.r, coll.cc.cmd[i], coll.cc.id[i]$. The latter are exactly the symbols associated with concrete values by the valuation functions $\sigma_i$. Therefore, it can be checked whether $\pi'$ is a model for $\Phi$ and consistent with the condition-action executions specified in the scenario type of $S$. Since we are dealing with finite sequences $\pi'$ only, a \textit{finite} interpretation of LTL semantics is required. For our purposes, the semantics proposed by De Giacomo and Vardi~\cite{DBLP:conf/ijcai/GiacomoV13} is most appropriate. We only need to adapt the Next-operator semantics to take the cycle time of objects into account. Let $\pi = \sigma_1\dots\sigma_k$ and assume that $1\le i\le k$. Then the following rules specify whether an LTL formula holds in position $i$ of $\pi'$.
\begin{enumerate}
\item $\pi',i \models \mathtt{expr}$ iff $\sigma_i \models \mathtt{expr}$. Here, $\mathtt{expr}$ is a Boolean expression over actual parameters and auxiliary variables  of the scenario instance.

\item $\pi',i \models \neg\varphi$ iff $\pi',i \not\models \varphi$.

\item $\pi',i \models \varphi_1\wedge \varphi_2$ iff $\pi',i \models \varphi_1$ and $\pi',i \models \varphi_2$.

\item Let $\varphi$ be an LTL formula referring to object parameters, parameters of $S$, and auxiliary variables of   $S$. Let $c \ge 1$ be the maximal cycle time of these objects.\footnote{If only parameters and auxiliary variables of $S$ are referenced in $\varphi$, $c = 1$, since elementary scenario instances are executed with maximal processing speed.}
Then 
$$\pi',i \models \tx \varphi\ \text{iff}\ \exists \ell\in \{ i+1,\dots, \min(k,2c-1) \}\centerdot \pi',\ell \models \varphi.$$

\item $\pi',i \models \varphi_1\tu \varphi_2$ iff there exists $i \le j \le k$ such that $\pi',j\models \varphi_2$ and $\forall \ell\in\{i,\dots,k-1\}\centerdot \pi',\ell\models \varphi_1$.

\item A formula holds on the complete sequence $\pi'$, if it holds at    the first index $i=1$, that is, $\pi' \models \varphi$ iff $\pi',1\models\varphi$.
\end{enumerate}
The usual syntactic abbreviations $\tf\varphi \equiv \ttt\tu\varphi$, 
$\tg \varphi \equiv \neg\tf \neg\varphi$, $\varphi_1\Rightarrow\varphi_2 \equiv \neg (\varphi_1\wedge\neg\varphi_2)$, and $\varphi_1\Leftrightarrow\varphi_2 \equiv (\varphi_1\Rightarrow\varphi_2\wedge \varphi_2\Rightarrow\varphi_1)$ are defined just as in conventional LTL.

The modified semantics of the Next-operator is motivated as follows. If the formula application is performed in step $i$, then $\varphi$ must become $\ttt$ in one of the next steps $i+1,\dots,i+2c-1$, since in the worst case, the slowest object needs $c-1$ steps to become aware of the state $\sigma_i$, and then another $c$ steps to perform the actions that will make $\psi$ evaluate to $\ttt$.

Note that in this finite-trace semantics $\tf \varphi$ must become true on $\pi'$. This is reasonable for elementary scenarios implementing test oracles: It is required to observe that $\varphi$ becomes $\ttt$ at least once, and the test execution must be long enough to make this happen. Conversely, $\tg\varphi$ just requires that $\varphi$ is not violated \textit{while $S$ is active}. This is -- of course -- not a proof that $\tg\varphi$ holds on infinite executions, but it reflects the state of practice to test safety properties with a finite test execution.

During the execution of $S$, all parameter symbols contained in the $\frm$-variable may be set by $S$, so that the LTL specifications become true or remain true. For observer scenario instances, $\frm$ is always empty, and the  consistency   of $\pi'$ with the LTL specifications must be ensured by  the referenced SUT objects alone.

For the effect of condition-action executions, the following rules apply.
\begin{enumerate}
\item If the precondition of $S$ evaluates to $\ttt$ for the first time in $\sigma_i$, then the  effect of the initial action becomes visible in $\sigma_{i+1}$.
\item If a condition-action has a guard condition $[g]$ that evaluates to $\ttt$ in $\sigma_j$ and $S$ is active in $\sigma_j$ and $\sigma_{j+1}$, then the effect of the associated action becomes visible in $\sigma_{j+1}$.

\item If a condition-action has a change condition $\chgc{\psi}$ and $\psi$ evaluates to $\fff$ in  $\sigma_{j-1}$ and to $\ttt$ in $\sigma_j$, then the  effect of the associated action becomes visible in $\sigma_{j+1}$, provided that $S$ is active in steps $(j-1),j,(j+1)$.

\item The action semantics is that of conventional while-languages~\cite{apt2010}, but with terminating loops only. With this interpretation, auxiliary variables that are not affected by any action in step $j$ remain unchanged in $\sigma_{j+1}$.

\end{enumerate}

\end{document}